\documentclass[lettersize,journal]{IEEEtran}
\usepackage{amsmath,amsfonts}
\usepackage{amsthm}
\usepackage{amssymb}
\usepackage{mathtools}

\usepackage{algorithm}
\usepackage{subcaption}
\usepackage{algpseudocode}

\usepackage{array}

\usepackage{textcomp}
\usepackage{stfloats}
\usepackage{url}
\usepackage{verbatim}
\usepackage{graphicx}
\usepackage{cite}
\usepackage{cuted}
\newtheorem{lemma}{Lemma} 
\def\BibTeX{{\rm B\kern-.05em{\sc i\kern-.025em b}\kern-.08em
    T\kern-.1667em\lower.7ex\hbox{E}\kern-.125emX}}

\begin{document}

\title{{\fontsize{24pt}{22pt}\selectfont Robust and Secure Transmission for Movable-RIS \\ Assisted ISAC with Imperfect Sense Estimation}}
        
\author{Ling Zhuang,\IEEEmembership{~IEEE Student Member,} Ximing Xie,\IEEEmembership{~IEEE Member,} Fang Fang,\IEEEmembership{~IEEE Senior Member,} \\ Ali Attaran, and Zhizhong Zhang\IEEEmembership{~IEEE Senior Member}
 \thanks{Ling Zhuang and Zhizhong Zhang are with the School of Electronic and Information Engineering, Nanjing University of Information Science and Technology, Nanjing, 210044, China (e-mail: 20211118008@nuist.edu.cn; zhangzz@nuist.edu.cn).
 
 Ximing Xie and Ali Attaran are with the Department of Electrical and Computer Engineering, Western University, London, ON, N6A 3K7, Canada (e-mail: xxie269@uwo.ca; aattaran@uwo.ca).
 
 Fang Fang is with the Department of Electrical and Computer Engineering and the Department of Computer Science, Western University, London, ON, N6A 3K7, Canada (e-mail: fang.fang@uwo.ca).}
}



\maketitle

\begin{abstract}
 Reconfigurable intelligent surfaces (RISs) have been extensively applied in integrated sensing and communication (ISAC) systems due to the capability of enhancing physical layer security (PLS). However, conventional static RIS architectures lack the flexibility required for adaptive beam control in multi-user and multifunctional scenarios. To address this issue without introducing additional hardware complexity and power consumption, in this paper, we exploit a movable RIS (MRIS) architecture, which consists of a large fixed sub-surface and a smaller movable sub-surface that slides on the fixed sub-surface to achieve dynamic beam reconfiguration with static phase shifts. This paper investigates an MRIS-assisted ISAC system under imperfect sensing estimation, where dedicated radar signals serve as artificial noise to enhance secure transmission against potential eavesdroppers (Eves). The transmit beamforming vectors, MRIS phase shifts, and relative positions of the two sub-surfaces are jointly optimized to maximize the minimum secrecy rate, ensuring robust secrecy performance for the weakest user under the uncertainty of the Eves’ channels. To handle the non-convexity, a convex bound is derived for the Eve channel uncertainty, and the $\mathcal{S}$-procedure is employed to reformulate semi-infinite constraints as linear matrix inequalities. An efficient alternating optimization and penalty dual decomposition-based algorithm is developed. Simulation results demonstrate that the proposed MRIS architecture substantially improves secrecy performance, especially when only a small number of elements are allocated to the movable sub-surface.
\end{abstract}

\section{Introduction}
\IEEEPARstart{W}{ith} the rapid evolution of wireless communication networks and the increasing scarcity of spectrum resources, information security has become increasingly critical \cite{r1}. The technology of physical layer security (PLS) directly protects information transmission by exploiting the inherent randomness and fading characteristics of wireless channels \cite{rnew1} and has attracted extensive attention. Recently, integrated sensing and communications (ISAC) has been proposed to unify communication and radar functions to improve spectral efficiency and channel state information (CSI) estimation \cite{r2, rr1}. Due to the sensing capability, ISAC enables real-time detection and localization of potential eavesdroppers (Eves), which can achieve adaptive security strategies that effectively overcome the inherent limitations of traditional ``blind design" approaches in PLS \cite{rr2}. However, ISAC systems have some challenges in limited sensing accuracy and coverage blind spots, particularly in complex propagation environments such as urban canyons, where severe wireless channel blockage and multipath fading further deteriorate system performance.


Reconfigurable intelligent surfaces (RISs) have emerged as a key technology for enabling virtual line-of-sight (LoS) links and intelligent signal reflection \cite{r3}. For example, RISs adaptively reconfigure beam patterns to enhance Eve localization accuracy in secure ISAC systems \cite{rr3}. Besides conventional RISs, advanced architectures such as simultaneously transmitting and reflecting RIS (STAR-RIS) \cite{r8}, stacked intelligent metasurfaces (SIM) \cite{r9}, and movable antenna RIS (MA-RIS) \cite{r10} further enhance PLS in ISCA systems. However, these RISs still face challenges in dynamic multi-user, multi-target scenarios requiring secure links, real-time sensing, and flexible beam control due to the lack of adaptability \cite{rnew2}, heavy control overhead \cite{r14}, and high operational complexity \cite{r15}.

To overcome these challenges, an innovative movable reconfigurable intelligent surface (MRIS) architecture has been proposed in \cite{r17}. MRIS is a dual-layer stacked structure consisting of a large, fixed base surface (S1) and a smaller, movable surface (S2) mounted on precision sliding tracks. With motor-driven control, S2 can be dynamically repositioned over the S1 surface. When S2 overlaps with S1 at different positions, it allows for the synthesis of diverse beam patterns. In particular, an MRIS prototype with the proof-of-concept experiments validated its capability to realize dynamic beam steering via differential position shifting of the S2 \cite{r17}. This mechanically adjustable configuration significantly enhances signal quality and extends coverage range, offering a cost-effective solution for dynamic beamforming.
\vspace{-4mm}
\subsection{Related Works}
In secure ISAC systems, RISs simultaneously enhance the received signal strength of legitimate users and suppress information leakage toward potential Eves, while enabling waveform optimization to improve sensing accuracy \cite{r4}. Recent studies have confirmed the effectiveness of RISs in enhancing PLS within ISAC frameworks.
For example, in \cite{r6}, an active RIS transceiver was proposed for secure ISAC, where rate-splitting multiple access (RSMA) and artificial noise (AN) were jointly employed to combat eavesdropping under imperfect CSI. A sum secrecy rate maximization problem was formulated and solved using a block coordinate descent-based approximation, which achieved joint optimization of transmit beamforming and time allocation. This approach improved both secrecy and energy efficiency while reducing the Cramér–Rao bound. Similarly, \cite{r5} investigated beamforming design for active and passive RIS-assisted ISAC systems, aiming to maximize beam gain subject to SINR and information leakage constraints. By employing a successive convex approximation (SCA) framework that jointly updates all design variables, the authors demonstrated significant gains in secrecy performance. They confirmed the superiority of active RISs over passive ones. Nevertheless, the performance advantages of active RISs come at the cost of greater hardware complexity, a higher computational burden, and longer solution time. Moreover, these studies are confined to static or quasi-static environments, limiting their applicability in dynamic networks. To extend RIS-assisted ISAC to time-varying scenarios, \cite{r7} developed a hybrid active–passive RIS-enhanced ISAC architecture for vehicle-to-everything networks, where the sum secrecy rate was maximized by jointly optimizing transmit beamforming, spectrum allocation, and RIS configurations via a hierarchical twin-delayed deep deterministic policy gradient algorithm. The hybrid RIS exhibited strong anti-eavesdropping capability and effectively mitigated channel degradation caused by vehicle mobility, thereby highlighting its potential for secure ISAC in dynamic environments. These advances establish a foundation for exploiting the spatial degrees of freedom (DoF) offered by RISs to enhance secure ISAC performance.

In addition to conventional architectures, several RIS variants have been employed to further strengthen secure ISAC. For example, \cite{r11} proposed an active STAR-RIS-assisted framework that integrates joint secrecy and covert communication, enabling full-space multiuser communication and half-space target sensing. In \cite{r12}, a hybrid active–passive SIM configuration was adopted to suppress interference by maximizing the worst-case achievable rate. Moreover, the MA-RIS achieves adaptive beam steering through dynamic antenna position reconfiguration, thereby enhancing integrated communication–sensing security performance while maintaining low RF chain complexity \cite{r13}.  

However, these RIS variants still have limited adaptability and control efficiency in dynamic multi-user and multi-target scenarios. Since the MRIS architecture was introduced in \cite{r17}, it has gained attention as a low-cost solution for dynamic beam control. In this case, \cite{rnew10} examined an MRIS-assisted sensing system and solved a worst-case sensing SINR maximization problem to improve multi-target sensing performance.

\vspace{-4mm}
\subsection{Motivation and Contributions}
The mechanical DoF introduced by MRIS enables flexible beam manipulation and spatial adaptability, which can be exploited to strengthen PLS by dynamically steering reflected energy away from Eves and enhancing spatial diversity for secure transmission. These properties make MRIS a promising architecture for secure ISAC systems. However, to the best of our knowledge, the integration of MRIS into secure ISAC remains unexplored. Meanwhile, most existing RIS-assisted secure ISAC designs assume perfect or partially known CSI of Eves, which is unrealistic since estimation errors are inevitable and can degrade secrecy performance \cite{rnew3}. Although the authors in \cite{r18} proposed a robust security scheme that characterized the uncertainty region of Eve’s CSI, their work was limited to single-target scenarios and ignored multipath fading effects. To overcome these limitations, this paper proposes a robust MRIS-assisted secure ISAC framework that explicitly accounts for multi-dimensional CSI uncertainties, including ranging errors, angular errors, and non-line-of-sight (NLoS) multipath scattering, arising from imperfect ISAC sensing. By deriving convex outer bounds for the non-convex uncertainty regions, we enable tractable robust optimization that ensures both communication secrecy and sensing accuracy in multi-user multi-target scenarios.

The main contributions of this paper are summarized as follows: 
\begin{itemize}
\item  We investigate a robust and secure transmission scheme for an MRIS-assisted ISAC system, where the dedicated sensing signal simultaneously serves as the AN. Considering inevitable estimation errors, the system can only rely on distance and angle values with accuracy deviations, while multipath fading errors are constrained within fixed ranges. We formulate a max–min secrecy-rate optimization problem to ensure robustness for the weakest legitimate user, jointly optimizing the BS transmit beamformer, AN, MRIS phase shifts, and the beam-pattern assignment.

\item  Under the Rician fading model, we derive tractable bounds based on the safe approximation theory to precisely characterize large-scale fading uncertainty caused by Eve's distance estimation errors and small-scale fading uncertainty arising from NLoS scattering randomness and angle estimation errors. The $\mathcal{S}$-procedure is employed to transform semi-infinite constraints into a finite number of linear matrix inequalities (LMIs).

\item  To tackle the non-convex problem, we first reformulate it by the weighted minimum mean square error (WMMSE) method to handle the max-min structure. Within the alternating optimization (AO) framework, the BS transmit beamforming subproblem is transformed into semidefinite programming subject to LMIs and quadratic constraints. Then, the beam pattern assignment problem, which is intrinsically coupled with inter-user interference, is reformulated using the big-M technique and strong Lagrangian duality to enable a tractable reformulation. Finally, the MRIS phase shift matrices are optimized independently by employing the penalty dual decomposition (PDD) method.

\item  Numerical results validate the effectiveness of the proposed robust scheme in achieving superior secrecy performance. In particular, the MRIS yields substantial secrecy gains even with a small mechanical displacement S2 compared with the conventional static RIS (SRIS), which indicates that the introduced mechanical DoF plays a crucial role in enhancing PLS within ISAC systems. 

\end{itemize}
\subsection{Organization}
This paper is organized as follows. We analyze the system model and formulate the optimization problem of the MRIS-assisted robust and secure ISAC systems in Section II.  In Section III, the bounds for Eve's CSI uncertainty regions and the proposed algorithm are presented. Simulation results are given in Section IV. Finally, conclusions are provided in Section V.

\textit{Notation:} We use $\mathbf{I}_{n}$ for the $n \times n$ identity matrix and $\mathrm{diag}(\mathbf{x})$ for a diagonal matrix with vector $\mathbf{x}$ on the diagonal. $\mathbb{C}^{a \times b}$ denotes the set of $a \times b$ complex matrices. The superscripts $(\cdot)^{H}$, $(\cdot)^{T}$, and $(\cdot)^{*}$ represent conjugate transpose, transpose, and conjugate, respectively. $|\cdot|$ and $\|\cdot\|$ denote the absolute value and Euclidean norm, respectively. $x \sim \mathcal{CN}(\mu, \sigma^2)$ indicates a complex Gaussian variable with mean $\mu$ and variance $\sigma^2$. $\Re(\cdot)$ extracts the real part, $\odot$ denotes Hadamard product, $\otimes$  denotes Kronecker product, $\oplus$ denotes Minkowski sum, and $[a]^{+} \triangleq \max\{0, a\}$.


\begin{figure}[!t]
\centering
\includegraphics[width=3in, trim=0 0 0 0, clip]{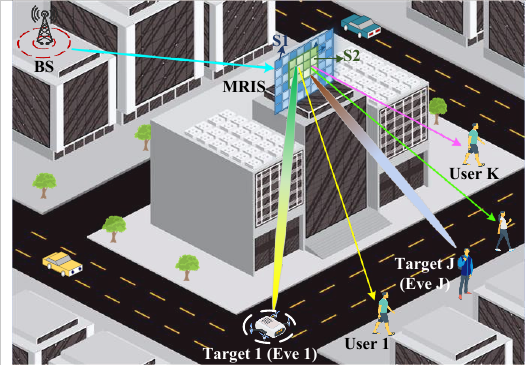}
\caption{The MRIS-assisted secure ISAC system.}
\label{fig1}
\end{figure}

\section{System Model and Problem Formulation}
As depicted in Fig. \ref{fig1}, an ISAC BS equipped with $L$ transmit antennas and $L$ receive antennas simultaneously serves $K$ single-antenna legitimate users and senses $J$ targets, which also act as potential Eves, assisted by an MRIS. We assume severe blockage conditions where direct BS-to-receiver links 
are negligible. The MRIS comprises a larger fixed base S1 and a smaller movable S2.  Specifically, S1 comprises $M=M_{\mathrm{r}}\times M_{\mathrm{c}}$ uniformly spaced transmissive elements arranged in $M_{\mathrm{r}}$ rows and $M_{\mathrm{c}}$ columns, while S2 has a similar arrangement with $N=N_{\mathrm{r}}\times N_{\mathrm{c}}$ elements. Elements on both surfaces are uniformly spaced at $d_{\mathrm{R}}=\lambda/\text{4}$ with $\lambda$ being the carrier wavelength. The elements of S1 and S2 are indexed respectively as $m=(m_{\mathrm{r}}-\text{1})M_{\mathrm{c}}+m_{\mathrm{c}}$ and $n=(n_{\mathrm{r}}-\text{1})N_{\mathrm{c}}+n_{\mathrm{c}}$, where $m_{\mathrm{r}}\in\mathcal{}\{\text{1},\dots,M_{\mathrm{r}}\}$, $m_{\mathrm{c}}\in\{\text{1},\dots,M_{\mathrm{c}}\}$, and similarly for $n_{\mathrm{r}}$ and $n_{\mathrm{c}}$. The phase shift matrices of S1 and S2 are denoted by $\boldsymbol{\Theta}=\mathrm{diag}(\boldsymbol{\theta})\in\mathbb{C}^{M\times M}$ and $\boldsymbol{\Phi}=\mathrm{diag}(\boldsymbol{\phi})\in\mathbb{C}^{N\times N}$, respectively. The corresponding phase shift vectors are $\boldsymbol{\theta}=[e^{j\theta_\text{1}},\dots,e^{j\theta_M}]^{T}$ and $\boldsymbol{\phi}=[e^{j\phi_\text{1}},\dots,e^{j\phi_N}]^{T}$. By sliding S2 over S1, the MRIS creates $B=B_{\mathrm{r}}\times B_{\mathrm{c}}$ distinct beam patterns, where $B_{\text{r}}=M_{\mathrm{r}}-N_{\mathrm{r}}+\text{1}$ and $B_{\mathrm{c}}=M_{\mathrm{c}}-N_{\mathrm{c}}+\text{1}$. Each beam pattern $b\in\mathcal{B}=\{\text{1},\text{2},\dots,B\}$ corresponds to a unique overlap configuration of S2 on S1. The equivalent phase shift vector for S2 at beam pattern $b$ is defined as $\bar{\boldsymbol{\phi}}_{b}=\boldsymbol{E}_{b}\boldsymbol{\phi}+\boldsymbol{e}_{b}\in\mathbb{C}^{M\times1},\forall b\in\mathcal{B}$, where the entries of $\boldsymbol{E}_{b}$ and $\boldsymbol{e}_{b}$ are given by 
 \begin{equation} 
 \left[\boldsymbol{E}_{b}\right]_{m,n}=\begin{cases}
 1, & \begin{array}{l}
 \textrm{if \ensuremath{n}-th element of S2 locates upon}\\
 \textrm{\ensuremath{m}-th element of S1},
 \end{array}\\
 \text{0}, & \begin{array}{c}
 \textrm{otherwise},\end{array}
 \end{cases} \label{6}
 \end{equation}
  \vspace*{0.45mm}
 \begin{equation} 
\left[\boldsymbol{e}_{b}\right]_{m}=\begin{cases}
 \text{0}, & \begin{array}{l}
 \textrm{if \ensuremath{n}-th element of S2 locates upon}\\
 \textrm{\ensuremath{m}-th element of S1},
 \end{array}\\
 1, & \begin{array}{c}
 \textrm{otherwise}.\end{array}
 \end{cases} \label{7}
 \end{equation}
 
 To characterize the beam pattern assignment, we denote $\boldsymbol{\chi}\in\left\{ \text{0},\text{1}\right\} ^{K \times B}$ as the binary beam pattern scheduling matrix, where $\chi_{k,b}=\text{1}$ indicates that the user $k$ is assigned to $b$-th beam pattern, and  $\sum_{b=\text{1}}^{B}\chi_{k,b}=\text{1}$  guarantees that the $k$-th user is exclusively assigned one beam pattern from the set $\mathcal{B}$.
 
 The transmit signal from the BS is expressed as $\mathbf{x}=\sum_{k=\text{1}}^{K}\boldsymbol{w}_{k}s_{k}+\mathbf{f}$, where $\boldsymbol{w}_{k}\in\mathbb{C}^{L\times \text{1}}$ is the communication beamforming vector and $s_{k}$ is the normalized communication symbol with $\mathbb{E}\{|s_{k}|^{\text{2}}\}=\text{1}$. The AN vector $\mathbf{f}\in\mathbb{C}^{L\times1}$ is designed for secure transmission and target sensing, following a complex Gaussian distribution $\mathbf{f}\sim\mathcal{CN}(\mathbf{0},\mathbf{F})$ with covariance matrix $\mathbf{F}\succeq\mathbf{0}$.  
 
 Considering the inevitable scattering in complex environments such as urban canyons, we adopt the Rician fading model to characterize  the  BS-to-MRIS channel $\boldsymbol{G}\in\mathbb{C}^{M\times L}$ as follows
 \begin{equation}
 \boldsymbol{G}=\frac{\sqrt{\beta_0}}{d_{\mathrm{BR}}}\left(\sqrt{\frac{\kappa_{\mathrm{BR}}}{\text{1}+\kappa_{\mathrm{BR}}}}\boldsymbol{G}^{\mathrm{LoS}}+\sqrt{\frac{\text{1}}{\text{1}+\kappa_{\mathrm{BR}}}}\boldsymbol{G}^{\mathrm{NLoS}}\right),
 \end{equation}
 where $\beta_0$ denotes the unit-distance path loss, $d_{\mathrm{BR}}$ is the distance between the BS and MRIS, and $\kappa_{\mathrm{BR}}$ is the Rician factor. The LoS component $\boldsymbol{G}^{\mathrm{LoS}}$ is defined accordingly based on steering vectors, i.e., 
 \begin{equation}
 \boldsymbol{G}^{\mathrm{LoS}}=\boldsymbol{a}_{\mathrm{MS}}\left(\vartheta_{\mathrm{BR}},\varphi_{\mathrm{BR}}\right)\boldsymbol{a}_{\mathrm{BS}}^{H}\left(\vartheta_{\mathrm{t}}\right), 
\end{equation} 
where $\vartheta_{\mathrm{t}}$, $\vartheta_{\mathrm{BR}}$, and $\varphi_{\mathrm{BR}}$ denote the angle of departure (AoD) at the BS, azimuth angle-of-arrival (AoA), and elevation AoA at the MRIS, respectively; $\boldsymbol{a}_{\mathrm{BS}}\left(\vartheta_{\mathrm{t}}\right)$ and  $\boldsymbol{a}_{\mathrm{MS}}\left(\vartheta_{\mathrm{BR}},\varphi_{\mathrm{BR}}\right)$ respectively denote the transmit array response at the BS and the receive array response at the MRIS, which are given by 
\begin{equation}
\boldsymbol{a}_{\mathrm{BS}}\left(\vartheta_{\mathrm{t}}\right)=[\text{1},e^{j\frac{\text{2}\pi d_{\mathrm{B}}}{\lambda}\sin\left(\vartheta_{\mathrm{t}}\right)},\cdots,e^{j\frac{\text{2}\pi d_{\mathrm{B}}}{\lambda}\left(L-\text{1}\right)\sin\left(\vartheta_{\mathrm{t}}\right)}]^{T}, \label{12}
\end{equation}
 \begin{equation}
\boldsymbol{a}_{\mathrm{MS}}\left(\vartheta_{\mathrm{BR}},\varphi_{\mathrm{BR}}\right)=[e^{j\text{2}\pi\delta_{\mathrm{r}}^{\mathrm{B}}\boldsymbol{m}_{\mathrm{r}}}]^{T}\otimes[e^{j\text{2}\pi\delta_{\mathrm{c}}^{\mathrm{B}}\boldsymbol{m}_{\mathrm{c}}}]^{T},
\end{equation}
with $\delta_{\mathrm{r}}^{\mathrm{B}}=\frac{d_{\mathrm{R}}}{\lambda}\cos\left(\vartheta_{\mathrm{BR}}\right)\sin\left(\varphi_{\mathrm{BR}}\right)$,
$\delta_{\mathrm{c}}^{\mathrm{B}}=\frac{d_{\mathrm{R}}}{\lambda}\sin\left(\vartheta_{\mathrm{BR}}\right)\sin\left(\varphi_{\mathrm{BR}}\right)$,
$\boldsymbol{m}_{\mathrm{r}}=[\text{0},\text{1},\cdots,M_{\mathrm{r}}-\text{1}]$, and $\boldsymbol{m}_{\mathrm{c}}=[\text{0},\text{1},\cdots,M_{\mathrm{c}}-\text{1}]$. The NLoS component $\boldsymbol{G}^{\mathrm{NLoS}}$ obeys the circularly symmetric complex Gaussian distribution with $\boldsymbol{G}^{\mathrm{NLoS}}\sim\mathcal{CN}\left(\boldsymbol{0},\mathbf{R}\right)$. 

Similarly, the MRIS-to-user channels, $\boldsymbol{h}_{\mathrm{U},k}\in\mathbb{C}^{M\times1}$, $k\in\mathcal{K}=\{\text{1},\text{2},\dots,K\}$, are modeled as Rician fading channels with LoS and NLoS components, which are given by 
\begin{equation}
\boldsymbol{h}_{\mathrm{U},k}	\!=\!\!\frac{\sqrt{\beta_{\text{0}}}}{d_{\mathrm{RU,}k}}\!\left(\!\sqrt{\frac{\kappa_{\mathrm{RU},k}}{\text{1}\!+\!\kappa_{\mathrm{RU},k}}}\boldsymbol{h}_{\mathrm{U},k}^{\mathrm{LoS}}+\!\right.
\left.\sqrt{\frac{\text{1}}{\text{1}\!+\!\kappa_{\mathrm{RU},k}}}\boldsymbol{h}_{\mathrm{U},k}^{\mathrm{NLoS}}\!\right)\!,   
\end{equation}
 where $d_{\mathrm{RU},k}$ denotes the distance between the MRIS and user $k$; the LoS component is $ \boldsymbol{h}_{\mathrm{U},k}^{\mathrm{LoS}}=\boldsymbol{a}_{\mathrm{MS}}\left(\vartheta_{\mathrm{RU},k},\varphi_{\mathrm{RU},k}\right)$, where $\vartheta_{\mathrm{RU},k}$ and $\varphi_{\mathrm{RU},k}$ are azimuth and elevation AoAs from the MRIS to users, respectively. Therefore, the received signal at the $k$-th user under beam pattern $b$ is expressed as
\begin{align}
y_{k,b}^{\mathrm{U}} & \!=\!\underset{\text{Desired signal}}{\underbrace{\boldsymbol{h}_{\mathrm{U},k}^{H}\bar{\boldsymbol{\Phi}}_{b}\boldsymbol{\Theta}\boldsymbol{G}\boldsymbol{w}_{k,b}s_{k,b}}}\!+\!\!\!\!\underset{\text{Multiuser interference on the same beam}}{\underbrace{\sum_{i\in\mathcal{K}\backslash\left\{ k\right\} }\!\!\!\chi_{i,b}\boldsymbol{h}_{\mathrm{U},k}^{H}\bar{\boldsymbol{\Phi}}_{b}\boldsymbol{\Theta}\boldsymbol{G}\boldsymbol{w}_{i,b}s_{i,b}}}\\
 & +\underset{\text{AN}}{\underbrace{\boldsymbol{h}_{\mathrm{U},k}^{H}\bar{\boldsymbol{\Phi}}_{b}\boldsymbol{\Theta}\boldsymbol{G}\mathbf{f}_{b}}}+n_{\mathrm{U},k},\nonumber 
\end{align}
 in which $n_{\mathrm{U},k}$ represents additive white Gaussian noise (AWGN) at the $k$-th user location and $\bar{\boldsymbol{\Phi}}_{b}=\mathrm{diag}(\bar{\boldsymbol{\phi}}_{b})$.
 Consequently, the achievable  rate of $k$-th user under beam pattern $b$ is $R_{k,b}^{\mathrm{U}}=\log\left(\text{1}+\mathrm{SINR}_{k,b}^{\mathrm{U}}\right)$, where 
\begin{equation}
\mathrm{SINR}_{k,b}^{\mathrm{U}}\!=\!\!\frac{\left|\left(\mathbf{h}_{k,b}^{\mathrm{U}}\right)^{H}\boldsymbol{w}_{k,b}\right|^{\text{2}}}{\!\!\underset{i\in\mathcal{K}\backslash\left\{ k\right\} }{\sum}\!\!\chi_{i,b}\left|\!\left(\mathbf{h}_{k,b}^{\mathrm{U}}\right)^{H}\boldsymbol{w}_{i,b}\right|^{\text{2}}\!\!\!+\!\left|\!\left(\mathbf{h}_{k,b}^{\mathrm{U}}\right)^{H}\mathbf{f}_{b}\right|^{\text{2}}\!\!\!+\!\sigma_{\mathrm{U},k}^{\text{2}}},
\end{equation}
and $\left(\mathbf{h}_{k,b}^{\mathrm{U}}\right)^{H}=\boldsymbol{h}_{\mathrm{U},k}^{H}\bar{\boldsymbol{\Phi}}_{b}\boldsymbol{\Theta}\boldsymbol{G}=\left(\bar{\boldsymbol{\phi}}_{b}\odot\boldsymbol{\theta}\right)^{T}\mathrm{diag}\left(\boldsymbol{h}_{\mathrm{U},k}^{H}\right)\boldsymbol{G}$.

 In the sensing stage, the BS receives echoes from potential Eves through MRIS-assisted refracted paths. The refracted signal received at the BS from the $j$-th Eve is expressed as
 \begin{equation}
y_{k,j,b}^{\mathrm{E}}=\boldsymbol{h}_{\mathrm{E},j}^{H}\bar{\boldsymbol{\Phi}}_{b}\boldsymbol{\Theta}\boldsymbol{G}\sum_{k\in\mathcal{K}}\chi_{k,b}\boldsymbol{w}_{k,b}s_{k,b}+\boldsymbol{h}_{\mathrm{E},j}^{H}\bar{\boldsymbol{\Phi}}_{b}\boldsymbol{\Theta}\boldsymbol{G}\mathbf{f}_{b}+n_{\mathrm{E},j},
\end{equation}
 in which $n_{\mathrm{E},j}$ represents AWGN at the $j$-th Eve location. The MRIS-to-Eve channels, $\boldsymbol{h}_{\mathrm{E},j}\in\mathbb{C}^{M\times1}$, $j\in\mathcal{J}=\left\{ \text{1},\text{2},\ldots, J\right\} $, are modeled as Rician fading channels with LoS and NLoS components, which are expressed as 
\begin{equation}
\boldsymbol{h}_{\mathrm{E},j}=\frac{\sqrt{\beta_{\text{0}}}}{d_{\mathrm{RE},j}}\left(\sqrt{\frac{\kappa_{\mathrm{RE},j}}{\text{1}+\kappa_{\mathrm{RE},j}}}\boldsymbol{h}_{\mathrm{E},j}^{\mathrm{LoS}}+\sqrt{\frac{\text{1}}{\text{1}+\kappa_{\mathrm{RE},j}}}\boldsymbol{h}_{\mathrm{E},j}^{\mathrm{NLoS}}\right),
\end{equation}
 where $d_{\mathrm{RE},j}$ denoting the distance between the MRIS and Eve $j$; the LoS component is   $\boldsymbol{h}_{\mathrm{E},j}^{\mathrm{LoS}}=\boldsymbol{a}_{\mathrm{MS}}\left(\vartheta_{\mathrm{RE},j},\varphi_{\mathrm{RE},j}\right)$, where $\vartheta_{\mathrm{RE},j}$ and $\varphi_{\mathrm{RE},j}$ are azimuth and elevation AoAs
from the MRIS to Eves, respectively.
 
In this work, we focus on the target-tracking phase of the ISAC process. It assumes that the target’s distance and angle have been estimated during the preceding detection stage \cite{rnew10}. However, due to hardware imperfections, estimation errors, and target dynamics, these estimates are inevitably inaccurate. Moreover, since the targets act as passive Eves, they do not cooperate with the BS to refine or correct their channel information. Specifically, we model the distance between the MRIS and the Eve $j$ as $\hat{d}_{\mathrm{RE},j}=\bar{d}_{\mathrm{RE},j}+\triangle d_{\mathrm{RE},j}$,
where $\bar{d}_{\mathrm{RE},j}$ is the nominal estimate and the distance uncertainty is bounded by $D_{\mathrm{RE},j}$, i.e., $\left|\triangle d_{\mathrm{RE},j}\right|\leq D_{\mathrm{RE},j}$.
Similarly, the azimuth and elevation are $\hat{\vartheta}_{\mathrm{RE},j}=\bar{\vartheta}_{\mathrm{RE},j}+\triangle\vartheta_{\mathrm{RE},j}$ and $ \hat{\varphi}_{\mathrm{RE},j}=\bar{\varphi}_{\mathrm{RE},j}+\triangle\varphi_{\mathrm{RE},j}$, with angular uncertainties bounded by $\varTheta_{\mathrm{RE},j}$ and $\varPsi_{\mathrm{RE},j}$, respectively, 
i.e., $\left|\triangle\vartheta_{\mathrm{RE},j}\right|\leq\varTheta_{\mathrm{RE},j}$ and $\left|\triangle\varphi_{\mathrm{RE},j}\right|\leq\varPsi_{\mathrm{RE},j}$. Meanwhile, due to scatterers and multipath fading, the Eves’ NLoS channels are generally unknown. We assume the
elements of $\boldsymbol{h}_{\mathrm{\mathrm{E}},j}^{\mathrm{NLoS}}$ are bounded by an upper boundary \cite{r6}, i.e., $\left|\tilde{h}_{\mathrm{E},j,m}\right|\leq\epsilon_{\mathrm{E},j,m}$. For convenience, we consolidate these uncertainties into the set 
\begin{align}
\Xi_{j} \triangleq 
\Bigl\{ 
|\triangle d_{\mathrm{RE},j}|\le D_{\mathrm{RE},j},\;
|\triangle\vartheta_{\mathrm{RE},j}|\le\varTheta_{\mathrm{RE},j},\\
|\triangle\varphi_{\mathrm{RE},j}|\le\varPsi_{\mathrm{RE},j},\;
|\tilde{h}_{\mathrm{E},j,m}|\le\epsilon_{\mathrm{E},j,m}
\Bigr\}. \nonumber
\end{align}

In summary, the channel from the MRIS to Eve $j$ can be reformulated as $\hat{\boldsymbol{h}}_{\mathrm{E},j}=\frac{\sqrt{\beta_{\text{0}}}}{\hat{d}_{\mathrm{RE},j}}\left(\sqrt{\frac{\kappa_{\mathrm{RE},j}}{\text{1}+\kappa_{\mathrm{RE},j}}}\hat{\boldsymbol{h}}_{\mathrm{E},j}^{\mathrm{LoS}}+\sqrt{\frac{\text{1}}{\text{1}+\kappa_{\mathrm{RE},j}}}\boldsymbol{h}_{\mathrm{E},j}^{\mathrm{NLoS}}\right)$ with $\hat{\boldsymbol{h}}_{\mathrm{E},j}^{\mathrm{LoS}}=\frac{\sqrt{\beta_{\text{0}}}}{\bar{d}_{\mathrm{RE},j}+\triangle d_{\mathrm{RE},j}}\boldsymbol{a}_{\mathrm{MS}}\left(\bar{\vartheta}_{\mathrm{RE},j}+\triangle\vartheta_{\mathrm{RE},j},\bar{\varphi}_{\mathrm{RE},j}+\triangle\varphi_{\mathrm{RE},j}\right)$.

 Then, the received SINR at the $j$-th Eve eavesdropping on the $k$-th user under the $b$-th beam pattern is $R_{_{k,j,b}}^{\mathrm{E}}=\log\left(\text{1}+\mathrm{SINR}_{_{k,j,b}}^{\mathrm{E}}\right)$, where 
 \begin{equation}
\mathrm{SINR}_{_{k,j,b}}^{\mathrm{E}}\!=\!\frac{\left|\left(\mathbf{h}_{j,b}^{\mathrm{E}}\right)^{H}\boldsymbol{w}_{k,b}\right|^{\text{2}}}{\underset{i\in\mathcal{K}\backslash\left\{ k\right\} }{\sum}\!\chi_{i,b}\left|\left(\mathbf{h}_{j,b}^{\mathrm{E}}\right)^{H}\!\boldsymbol{w}_{i,b}\right|^{\text{2}}\!\!+\!\left|\left(\mathbf{h}_{j,b}^{\mathrm{E}}\right)^{H}\!\mathbf{f}_{b}\right|^{\text{2}}\!\!+\!\sigma_{\mathrm{E},j}^{\text{2}}}, 
\end{equation}
and $\left(\mathbf{h}_{j,b}^{\mathrm{E}}\right)^{H}=\hat{\boldsymbol{h}}_{\mathrm{E},j}^{H}\bar{\boldsymbol{\Phi}}_{b}\boldsymbol{\Theta}\boldsymbol{G}$. 

 The target detection performance can be improved by enhancing the beam pattern gain in the Eve direction, which is given by
\begin{align}
\mathcal{P}_{j,b} & =\left\{ \left|\left(\mathbf{h}_{j,b}^{\mathrm{E}}\right)^{H}\mathbf{x}_{b}\right|^{\text{2}}\right\} \\
 & =\left(\mathbf{h}_{j,b}^{\mathrm{E}}\right)^{H}\left(\sum_{k=\text{1}}^{K}\chi_{k,b}\boldsymbol{w}_{k,b}\boldsymbol{w}_{k,b}^{H}+\mathbf{f}_{b}\mathbf{f}_{b}^{H}\right)\mathbf{h}_{j,b}^{\mathrm{E}}\nonumber.
\end{align}

 The corresponding achievable secrecy rate of the user $k$ on beam pattern $b$ is defined by $R_{k,b}^{\mathrm{S}}=\left[\text{0}, R_{k,b}^{\mathrm{U}}-\underset{\Xi_{j}}{\max}\, R_{k,j,b}^{\mathrm{E}}\right]^{+}$. Accordingly, the system aims to maximize the minimum secrecy rate by jointly optimizing the BS beamformers, AN, MRIS phases, and beam pattern assignment. The problem is formulated as
 \begin{align}
P0:\, & \underset{\left\{ \boldsymbol{W}_{b},\mathbf{f}_{b}\right\} _{b=\text{1}}^{B},\boldsymbol{\phi},\boldsymbol{\theta},\boldsymbol{\chi}}{\mathrm{max}}\enspace\underset{k\in\mathcal{K}}{\min}\enspace\sum_{b=\text{1}}^{B}\chi_{k,b}R_{k,b}^{\mathrm{S}}\\
 & \mathrm{s.t.}\,C\text{1}:\sum_{k=\text{1}}^{K}\chi_{k,b}\left\Vert \boldsymbol{w}_{k,b}\right\Vert ^{\text{2}}+\left\Vert \mathbf{f}_{b}\right\Vert ^{\text{2}}\leq P_{\mathrm{max}},\forall b,\nonumber \\
 & \enspace\quad C\text{2}:\sum_{b=\text{1}}^{B}\chi_{k,b}=\text{1},\forall k,\nonumber \\
 & \enspace\quad C\text{3}:\chi_{k,b}\in\left\{ \text{0},\text{1}\right\} ,\forall k,\forall b,\nonumber \\
 & \enspace\quad C\text{4}:\sum_{b=\text{1}}^{B}\chi_{k,b}R_{k,b}^{\mathrm{U}}\geq\Gamma_{k},\forall k,\nonumber \\
 & \enspace\quad C\text{5}:\mathcal{P}_{j,b}\geq\Gamma_{j}^{\mathrm{E}},\Xi_{j},\forall j,\forall b,\nonumber \\
 & \enspace\quad C\text{6}:\left|\theta_{m}\right|=\text{1},\forall m,\nonumber \\
 & \enspace\quad C\text{7}:\left|\phi_{n}\right|=\text{1},\forall n,\nonumber 
 \label{eq:P0}
\end{align}
where $\boldsymbol{W}_b{=}\left\{\boldsymbol{w}_{k,b}|\forall k\right\}$ are BS beamformers under $b$-th beam pattern.
$C\text{1}$ enforces the transmit power budget $P_{\max}$ for each beam pattern;
$C\text{2}$ and $C\text{3}$ ensure a unique binary assignment of pattern $b$ to user $k$; 
$C\text{4}$ guarantees the QoS of users; 
$C\text{5}$ satisfies sensing requirements under worst-case CSI uncertainty set $\Xi_j$; 
$C\text{6}$ and $C\text{7}$ impose unit-modulus MRIS phases on the two sub-surfaces. Problem $P$0 is nonconvex due to the coupled variables $\left\{\boldsymbol{W}_{b},\mathbf{f}_{b},\boldsymbol{\phi},\boldsymbol{\theta},\boldsymbol{\chi}\right\}$, bilinear couplings, worst-case terms, and unit-modulus constraints. To address this, we propose an AO-based algorithm that decomposes $P$0 into four tractable sub-problems solved iteratively.

\section{Proposed Algorithm}
To tackle the non-convex problem $P$0, we develop a three-stage solution 
framework: \textit{(i)} We first derive tractable bounds for the CSI uncertainty regions of Eve channels to enable robust optimization; \textit{(ii)} Then, we reformulate the intractable secrecy rate objective and CSI-dependent 
constraints into equivalent tractable forms via the WMMSE and $\mathcal{S}$-procedure, respectively; 
\textit{(iii)} Finally, we employ AO with PDD frameworks to obtain a suboptimal solution through iterative refinement of the four sub-problems.
\vspace{-2mm}
\subsection{ Bound for Eve's CSI Uncertainty Region}
In this subsection, we derive analytic geometry bounds on the uncertainty regions of each Eve's channel to enable robust and secure transmission. Under the Rician fading model, the analysis accounts for large-scale fading uncertainty primarily induced by distance estimation errors in path loss modeling, and small-scale fading uncertainty arising from NLoS scattering randomness and angular estimation errors affecting the LoS component. The $\hat{\boldsymbol{h}}_{\mathrm{E},j}$  can be rewritten as 
\begin{equation}
\hat{\boldsymbol{h}}_{\mathrm{E},j}\!=\!\hat{d}_{\mathrm{RE},j}^{-\text{1}}\left[\beta_{\text{1},j}\!+\!\beta_{\text{2},j}\tilde{h}_{\mathrm{E},j,\text{1}},\!\cdots,\!\beta_{\text{1},j}e^{j\hat{\psi}_{j,M}}\!+\!\beta_{\text{2},j}\tilde{h}_{\mathrm{E},j,M}\right]^{T}\!\!\!,
\end{equation}
where $\beta_{\text{1},j}=\sqrt{\frac{\beta_{\text{0}}\kappa_{\mathrm{RE},j}}{\left(\text{1}+\kappa_{\mathrm{RE},j}\right)}}$,
$\beta_{\text{2},j}=\sqrt{\frac{\beta_{\text{0}}}{\left(\text{1}+\kappa_{\mathrm{RE},j}\right)}}$, and 
\begin{equation}
\hat{\psi}_{j,m}=\text{2}\pi\left(\hat{\delta}_{\mathrm{r},j}^{\mathrm{E}}\left(m_{\mathrm{r}}-\text{1}\right)+\hat{\delta}_{\mathrm{c},j}^{\mathrm{E}}\left(m_{\mathrm{c}}-\text{1}\right)\right),
\end{equation}
with $\hat{\delta}_{\mathrm{r},j}^{\mathrm{E}}=\frac{d_{\mathrm{R}}}{\lambda}\cos\left(\hat{\vartheta}_{\mathrm{RE},j}\right)\sin\left(\hat{\varphi}_{\mathrm{RE},j}\right)$ and $\hat{\delta}_{\mathrm{c},j}^{\mathrm{E}}=\frac{d_{\mathrm{R}}}{\lambda}\sin\left(\hat{\vartheta}_{\mathrm{RE},j}\right)\sin\left(\hat{\varphi}_{\mathrm{RE},j}\right)$. 

To simplify notation, we omit the subscript $j$ in the following 
analysis. The LoS component $\beta_{1}\hat{d}_{\mathrm{RE}}^{-1}e^{j\hat{\psi}_{m}}$ 
represents a complex point in the complex plane with phase $\hat{\psi}_{m}$ 
and magnitude $R_{m}=\beta_{1}\left(\bar{d}_{\mathrm{RE}}+\triangle d_{\mathrm{RE}}\right)^{-1}$. 
Since distance estimation errors are bounded, the magnitude varies within 
the interval $R_{m}\in\left[R_{m,\inf},R_{m,\sup}\right]$, where
$R_{m,\inf}=\beta_{\text{1}}\left(\bar{d}_{\mathrm{RE}}+D_{\mathrm{RE}}\right)^{-\text{1}}$
and $R_{m,\sup}=\beta_{\text{1}}\left(\bar{d}_{\mathrm{RE}}-D_{\mathrm{RE}}\right)^{-\text{1}}$. Angular uncertainties $\bigtriangleup\vartheta_{\mathrm{RE}}$ and $\bigtriangleup\varphi_{\mathrm{RE}}$ introduce phase perturbations bounded by $\triangle\psi_{m}$, which is expressed as
\begin{equation}
\triangle\psi_{m}\!=\!\text{2}\pi\left|\left(\hat{\delta}_{\mathrm{r}}^{\mathrm{E}}\!-\!\bar{\delta}_{\mathrm{r}}^{\mathrm{E}}\right)\left(m_{\mathrm{r}}\!-\!\text{1}\right)\!+\!\left(\hat{\delta}_{\mathrm{c}}^{\mathrm{E}}\!-\!\bar{\delta}_{\mathrm{c}}^{\mathrm{E}}\right)\left(m_{\mathrm{c}}\!-\!\text{1}\right)\right|,
\end{equation}
where $\bar{\delta}_{\mathrm{r}}^{\mathrm{E}}=\frac{d_{\mathrm{R}}}{\lambda}\cos\left(\bar{\vartheta}_{\mathrm{RE}}\right)\sin\left(\bar{\varphi}_{\mathrm{RE}}\right)$ and $\bar{\delta}_{\mathrm{c}}^{\mathrm{E}}=\frac{d_{\mathrm{R}}}{\lambda}\sin\left(\bar{\vartheta}_{\mathrm{RE}}\right)\sin\left(\bar{\varphi}_{\mathrm{RE}}\right)$ represent the true spatial frequencies, respectively.

The NLoS component $\beta_{\text{2}}\left(\bar{d}_{\mathrm{RE}}+\bigtriangleup d_{\mathrm{RE}}\right)^{-\text{1}}\tilde{h}_{\mathrm{E},m}$ with bounded magnitude $\left|\tilde{h}_{\mathrm{E},m}\right|\leq\epsilon_{\mathrm{E},m}$ forms a circular uncertainty region centered on the LoS component with radius $r_{m}=\beta_{\text{2}}\left(\bar{d}_{\mathrm{RE}}+\bigtriangleup d_{\mathrm{RE}}\right)^{-\text{1}}\epsilon_{\mathrm{E},m}$. The distance error cause $r_{m}$ to vary within $r_{m}\in\left[r_{m,\inf},r_{m,\sup}\right]$, where $r_{m,\inf}=\beta_{\text{2}}\left(\bar{d}_{\mathrm{RE}}+D_{\mathrm{RE}}\right)^{-\text{1}}\epsilon_{\mathrm{E},m}$ and $r_{m,\sup}=\beta_{\text{2}}\left(\bar{d}_{\mathrm{RE}}-D_{\mathrm{RE}}\right)^{-\text{1}}\epsilon_{\mathrm{E},m}$.

\begin{figure}[!t]
\centering
\includegraphics[width=2.8in, trim=16 20 16 20, clip]{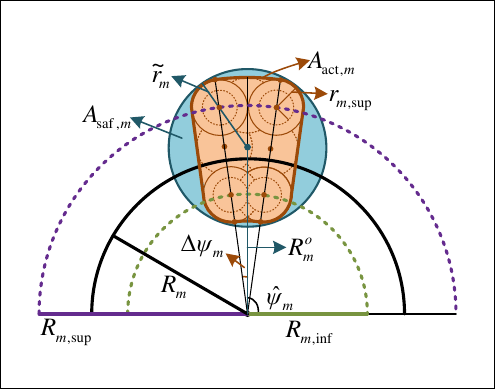}
\caption{ Geometric illustration of multi-dimensional CSI uncertainty bound.}
\label{fig2}
\vspace{-3mm}
\end{figure}

The combined effect of distance, angular, and multipath uncertainties creates a non-convex uncertainty region with circular arc boundaries, as depicted by the orange-colored area in Fig. \ref{fig2}. Using Minkowski sum operations \cite{rnew4}, the area of this region is
\begin{align}
A_{\mathrm{act},m}= & A_{\mathrm{det},m}\oplus A_{\mathrm{ran},m}\\
= & \pi r_{m,\sup}^{\text{2}}+\text{2}\left(R_{m,\sup}-R_{m,\inf}\right)r_{m,\sup}+ \nonumber \\
 & \left[\!\pi\left(R_{m,\sup}\!+\!\!r_{m,\sup}\right)^{\text{2}}\!\!-\!\!\pi\left(R_{m,\inf}\!-\!r_{m,\sup}\right)^{\text{2}}\!\right]\frac{\text{2}\triangle\psi_{m}}{\text{2}\pi}\nonumber \\
= & \pi r_{m,\sup}^{\text{2}}+\triangle\psi_{m}R_{m,\mathrm{sum}}\left(\text{2}r_{m,\sup}+\triangle R_{m,\max}\right)\nonumber \\
 & +\text{2}r_{m,\sup}\triangle R_{m,\max}.\nonumber
\end{align}
where $A_{\mathrm{det},m}$ represents the uncertainty region of the deterministic LoS component, $A_{\mathrm{ran},m}$ denotes the uncertainty region of the stochastic NLoS component, $\triangle R_{m,\max}=R_{m,\sup}-R_{m,\inf}$, and $R_{m,\mathrm{sum}}=R_{m,\sup}+R_{m,\inf}$.

The uncertainty region is non-convex, which complicates the robust beamforming design, as standard techniques such as the $\mathcal{S}$-procedure require convex uncertainty sets. To address this issue, we derive a convex outer bound by inscribing the uncertainty region within a circle, depicted
as the blue circle in Fig. \ref{fig2}. Based on geometric analysis, the radius of this circular bound is $\widetilde{r}_{m}=\sqrt{\left(R_{m}^{o}\right)^{2}+R_{m,\mathrm{out}}^{2}-2R_{m}^{o}R_{m,\mathrm{out}}\cos\left(\triangle\psi_{m}\right)}$, where the center magnitude is
$R_{m}^{o}=\frac{R_{m,\mathrm{out}}^{2}-R_{m,\mathrm{inn}}^{2}}{2\left(R_{m,\mathrm{out}}\cos\left(\triangle\psi_{m}\right)-R_{m,\mathrm{inn}}\right)}$, $R_{m,\mathrm{out}}=R_{m,\sup}+r_{m,\sup}$, and $R_{m,\mathrm{inn}}=R_{m,\inf}-r_{m,\sup}$.

The area of the circular outer bound is
\begin{equation}
A_{\mathrm{saf},m}=\pi\left(\widetilde{r}_{m}\right)^{\text{2}}.
\end{equation}

\begin{figure}[!t]
\centering
\includegraphics[width=3.4in, trim=0 0 0 0, clip]{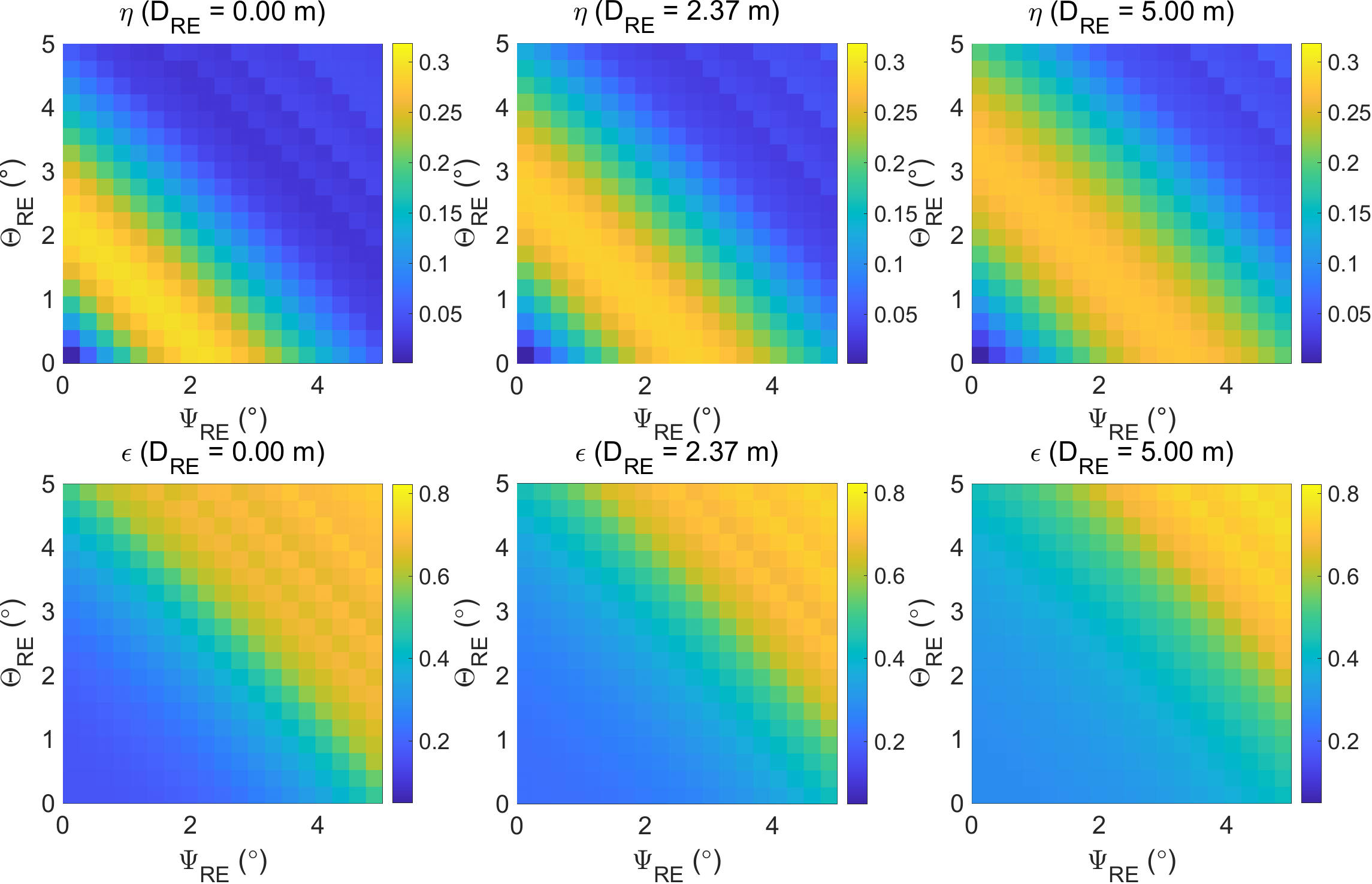}
\caption{Bound tightness $\eta$ (top) and channel error $\varepsilon$ 
(bottom) for different distances and phase errors with 
$\epsilon_{\mathrm{E},m}=0.377$.}
\label{fig3}
\vspace{-3mm}
\end{figure}

To validate the effectiveness of the proposed approximate bound, we present the bound tightness and channel error under the configuration $\bar{\vartheta}_{\mathrm{RE}}=\pi/3$, $\bar{\varphi}_{\mathrm{RE}}=\pi/3$, $\bar{d}_{\mathrm{RE}}=50$m, and 
$M=$10, as illustrated in Fig. \ref{fig3}. The bound tightness and normalized channel error are respectively quantified by metrics as follows 
\begin{equation}
\eta=\frac{\sum_{m=\text{1}}^{M}A_{\mathrm{act},m}}{\sum_{m=\text{1}}^{M}A_{\mathrm{saf},m}},
\end{equation}
\begin{equation}
\epsilon=\frac{\sqrt{\sum_{m=\text{1}}^{M}\left(\widetilde{r}_{m}\right)^{\text{2}}}}{\sqrt{\sum_{m=\text{1}}^{M}\left(R_{m}^{o}\right)^{\text{2}}}},
\end{equation} 
where $\eta$ represents the overall ratio of the actual uncertainty region area to the circumscribed bound area across all MRIS elements, and $\epsilon$ denotes the normalized channel error accounting for the radius deviations of individual antenna elements.

Fig. \ref{fig3} demonstrates the performance of the proposed uncertainty bound under varying parameter configurations. The top row presents the bound tightness $\eta$, while the bottom row illustrates the normalized channel error $\epsilon$. The distribution of $\eta$ confirms that the proposed bound effectively envelops the composite uncertainty region induced by the joint effects of distance, angular, and NLoS errors across the entire parameter space. A clear spatial gradient pattern emerges in $\eta$, such that as angular errors increase, the bound tightness $\eta$ also increases. This indicates that phase perturbations significantly affect the 
geometry of the actual uncertainty region. Although $\eta$ values remain relatively low, indicating a degree of conservativeness, this property ensures stable coverage of the actual uncertainty set by the circumscribed bound, which constitutes the necessary safety margin for robust optimization. The distribution of $\epsilon$ reveals a dual-impact mechanism of parameter uncertainties on channel error. Within each subplot, as the maximum angular errors $\varTheta_{\mathrm{RE}}$ and $\varPsi_{\mathrm{RE}}$ increase, $\epsilon$ escalates from approximately 0.1 to 0.8, primarily attributed to the accumulation of phase perturbations in the array manifold vector. Across subplots, as the distance error bound $D_{\mathrm{RE}}$ increases from 0 to 5m, the overall distribution of $\epsilon$ shifts from low-error regions toward high-error regions, demonstrating that distance uncertainty exerts a dominant influence on channel error through path loss attenuation and phase deviation. Notably, $\epsilon$ exceeds 0.5 only in high-angle and large-distance error regions, while remaining below 0.5 in a large portion of the parameter space. In summary, the proposed uncertainty bound construction method provides a tractable characterization of a convex set for subsequent $\mathcal{S}$-Procedure based robust optimization, ensuring theoretical security guarantees.

Then, we decompose the channel vector as $\boldsymbol{h}_{\mathrm{E},j}=\bar{\boldsymbol{h}}_{\mathrm{E},j}+\triangle\boldsymbol{h}_{\mathrm{E},j}$, where the nominal channel is
\begin{align}
\bar{\boldsymbol{h}}_{\mathrm{E},j}\triangleq & \left[R_{j,\text{1}}^{o},\cdots,R_{j,m}^{o}e^{j\text{2}\pi\left(\bar{\delta}_{\mathrm{r},j}^{\mathrm{E}}\left(m_{\mathrm{r}}-\text{1}\right)+\bar{\delta}_{\mathrm{c},j}^{\mathrm{E}}\left(m_{\mathrm{c}}-\text{1}\right)\right)},\cdots,\right.\\
 & \left.R_{j,M}^{o}e^{j\text{2}\pi\left(\bar{\delta}_{\mathrm{r},j}^{\mathrm{E}}\left(M_{\mathrm{r}}-\text{1}\right)+\bar{\delta}_{\mathrm{c},j}^{\mathrm{E}}\left(M_{\mathrm{c}}-\text{1}\right)\right)}\right]^{T}.\nonumber 
\end{align}

For tractable robust optimization, we define a spherical uncertainty set that bounds the channel error vector $\boldsymbol{\triangle}\boldsymbol{h}_{\mathrm{E},j}$ as
\begin{equation}
\boldsymbol{\triangle}\boldsymbol{h}_{\mathrm{E},j}^{H}\boldsymbol{\triangle}\boldsymbol{h}_{\mathrm{E},j}\leq\sum_{m=\text{1}}^{M}\left(\widetilde{r}_{j,m}\right)^{\text{2}}\triangleq\varepsilon_{j}^{\text{2}},
\end{equation}
\begin{equation}
\widetilde{\Xi}_{j}\triangleq\left\{ \boldsymbol{\triangle}\boldsymbol{h}_{\mathrm{E},j}\in\mathbb{C}^{M\times1}:\left\Vert \boldsymbol{\bigtriangleup}\boldsymbol{h}_{\mathrm{E},j}\right\Vert _{\text{2}}\leq\varepsilon_{j}\right\}.
\end{equation}

\vspace{-4mm}
\subsection{ Problem transformation}
In the following, we utilize the derived uncertainty bounds of potential Eves' channels to reformulate the original problem. By replacing the complex uncertainty set $\Xi_j$ with the tractable spherical bound $\widetilde{\Xi}_j$, we obtain the following reformulated problem
\begin{align}
P\text{1}:\, & \underset{\left\{ \boldsymbol{W}_{b},\mathbf{f}_{b}\right\} _{b=\text{1}}^{B},\boldsymbol{\phi},\boldsymbol{\theta},\boldsymbol{\chi}}{\mathrm{max}}\enspace\underset{k\in\mathcal{K}}{\min}\enspace\sum_{b=\text{1}}^{B}\chi_{k,b}\left(R_{k,b}^{\mathrm{U}}-\underset{\widetilde{\Xi}_{j}}{\max}\,R_{k,j,b}^{\mathrm{E}}\right)\\
 & \mathrm{s.t.}\,C\text{1}-C\text{4},C\text{6},C\text{7},\nonumber \\
 & \enspace\quad C\text{5}a:\mathcal{P}_{j,b}\geq\Gamma_{j}^{\mathrm{E}}, \widetilde{\Xi}_{j}, \forall j,\forall b,\nonumber 
\end{align}

Problem $P$1 remains challenging due to the non-convex objective function involving coupled variables $\{\boldsymbol{W}_{b}, \mathbf{f}_{b}, \boldsymbol{\phi}, \boldsymbol{\theta}, \boldsymbol{\chi}\}$ and the worst-case optimization over uncertainty sets. To address these challenges, we first transform the objective function using the WMMSE method \cite{r24}. By introducing the auxiliary variable $\boldsymbol{\mu}=\left[\mu_{\text{1},\text{1}},\cdots,\mu_{K,\text{1}};\cdots;\mu_{\text{1},B}\cdots,\mu_{K,B}\right]$, the corresponding MSE of user $k$ on beam pattern $b$ is expressed as
\begin{align}
& e_{k,b}\left(\boldsymbol{W}_{b},\mathbf{f}_{b},\boldsymbol{\phi},\boldsymbol{\theta},\boldsymbol{\chi},\mu_{k,b}\right)=\\
&  \left(\text{1}-\mu_{k,b}^{\ast}\left(\mathbf{h}_{k,b}^{\mathrm{U}}\right)^{H}\boldsymbol{w}_{k,b}\right)\left(\text{1}-\mu_{k,b}^{\ast}\left(\mathbf{h}_{k,b}^{\mathrm{U}}\right)^{H}\boldsymbol{w}_{k,b}\right)^{H}+ \nonumber\\
& \left|\mu_{k,b}\right|^{\text{2}}\left(\underset{i\in\mathcal{K}\backslash\left\{ k\right\} }{\sum}\chi_{i,b}\left|\left(\mathbf{h}_{k,b}^{\mathrm{U}}\right)^{H}\boldsymbol{w}_{i,b}\right|^{\text{2}}+\left|\left(\mathbf{h}_{k,b}^{\mathrm{U}}\right)^{H}\mathbf{f}_{b}\right|^{\text{2}}+\sigma_{\mathrm{U},k}^{\text{2}}\right).\nonumber
\end{align}

 Then, we invoke the weights $\boldsymbol{z}=\left[z_{\text{1},\text{1}},\cdots,z_{K,\text{1}};\cdots;z_{\text{1},B},\cdots,z_{K,B}\right]$ and the cost function $
\mathrm{log}\left(\text{1}+\mathrm{SINR}_{k,b}^{\mathrm{U}}\right)=\underset{\boldsymbol{z}\succ\boldsymbol{0},\boldsymbol{\mu}}{\max}\mathrm{log}\left(\boldsymbol{z}\right)-\mathrm{Tr}\left(\boldsymbol{z}e_{k,b}\right)+\text{1}$ for the transmit filters, the rate of the $k$-th user under beam pattern $b$ can be equivalently written as
\begin{align}
R_{k,b}^{\mathrm{U}}= & \underset{z_{k,b}\geq0,\mu_{k,b}}{\max}y_{k,b}\\
= & \underset{z_{k,b}^{\mathrm{U}}\geq0,\mu_{k,b}}{\max}\text{2}z_{k,b}\Re\left\{ \mu_{k,b}^{\ast}\left(\mathbf{h}_{k,b}^{\mathrm{U}}\right)^{H}\boldsymbol{w}_{k,b}\right\} +\widetilde{z}_{k,b}-\nonumber \\
 & z_{k,b}\left|\mu_{k,b}\right|^{\text{2}}\left(\underset{i\in\mathcal{K}}{\sum}\chi_{i,b}\left|\left(\mathbf{h}_{k,b}^{\mathrm{U}}\right)^{H}\boldsymbol{w}_{i,b}\right|^{\text{2}}+\left|\left(\mathbf{h}_{k,b}^{\mathrm{U}}\right)^{H}\mathbf{f}_{b}\right|^{\text{2}}\right),\nonumber 
\end{align}
where $\widetilde{z}_{k,b}=\log\left(z_{k,b}\right)-z_{k,b}-z_{k,b}\left|\mu_{k,b}\right|^{\text{2}}\sigma_{\mathrm{U},k}^{\text{2}}+\text{1}$.
 
To further simplify the objective function of $P$1, we introduce slack variables $t$, $\boldsymbol{v}=\left[v_{\text{1},\text{1}},\cdots,v_{K,\text{1}};\cdots;v_{\text{1},B},\cdots,v_{K,B}\right]^{T}$ and $\bar{\boldsymbol{v}}=\left[\bar{v}_{\text{1},\text{1},\text{1}},\cdots,\bar{v}_{\text{1},J,\text{1}};\cdots;\bar{v}_{K,\text{1},B}\cdots,\bar{v}_{K,J,B}\right]$ for Eves' rates. 
Then, problem $P$1 can be reformulated as 
\begin{align}
P\text{2}:\, & \underset{\begin{array}[t]{l}
\left\{ \boldsymbol{W}_{b},\mathbf{f}_{b}\right\} _{b=\text{1}}^{B},\boldsymbol{\phi},\\
\boldsymbol{\theta},\boldsymbol{\chi},t,\boldsymbol{z},\boldsymbol{\mu},\boldsymbol{v},\bar{\boldsymbol{v}}
\end{array}}{\mathrm{max}}\enspace t\\
 & \mathrm{s.t.}\,C\text{1}-C\text{3},C\text{5}a,C\text{6},C\text{7},\nonumber \\
 & \enspace\quad C\text{4}a:\sum_{b=\text{1}}^{B}\chi_{k,b}y_{k,b}\geq\Gamma_{k},\forall k,\nonumber \\
 & \enspace\quad C\text{8}:\sum_{b=\text{1}}^{B}\chi_{k,b}\left(y_{k,b}-v_{k,b}\right)\geq t,\forall k,\nonumber \\
 & \enspace\quad C\text{9}:\!\left|\left(\mathbf{h}_{j,b}^{\mathrm{E}}\right)^{H}\!\!\boldsymbol{w}_{k,b}\right|^{\text{2}}\!\leq\!\bar{v}_{k,j,b}\left(\text{2}^{v_{k,b}}\!\!-\!\!\text{1}\right)\!,\widetilde{\Xi}_{\mathrm{E},j},\!\forall k,\!\forall j,\!\forall b,\nonumber \\
 & \enspace\quad C\text{10}:\!\sum_{i\in\mathcal{K}\backslash\left\{ k\right\} }\chi_{i,b}\left|\left(\mathbf{h}_{j,b}^{\mathrm{E}}\right)^{H}\boldsymbol{w}_{i,b}\right|^{\text{2}}+\left|\left(\mathbf{h}_{j,b}^{\mathrm{E}}\right)^{H}\mathbf{f}_{b}\right|^{\text{2}}+\nonumber \\
 & \enspace\quad\enspace\quad\sigma_{\mathrm{E},j}^{\text{2}}\geq\bar{v}_{k,j,b},\widetilde{\Xi}_{\mathrm{E},j},\forall k,\forall j,\forall b.\nonumber
\end{align}
\subsection{Semi-Infinite Constraint Transformation}
 To handle the semi-infinite CSI uncertainty constraints $C\text{5}a$, $C\text{9}$, and $C\text{10}$, we employ the following generalized $\mathcal{S}$-procedure to transform them into tractable LMIs. 
 \begin{lemma}[$\mathcal{S}$-procedure \cite{r23}]
Let a quadratic
function $f_{i}\left(\mathbf{x}\right)$, $\mathbf{x}\in\mathbb{C}^{N\times\text{1}}$,
$i\in\mathcal{I}=\left\{ \text{0},\text{1},\cdots,I\right\}$, be defined as
\begin{equation}
f_{i}\left(\mathbf{x}\right)=\mathbf{x}^{H}\mathbf{A}_{i}\mathbf{x}+\text{2}\Re\left\{ \mathbf{a}_{i}^{H}\mathbf{x}\right\} +a_{i},
\end{equation}
where $\mathbf{A}_{i}\in\mathbb{H}^{N}$, $\mathbf{a}_{i}\in\mathbb{C}^{N\times\text{1}}$,
and $a_{i}\in\mathbb{R}$. Then the condition $\left\{ f_{i}\left(\mathbf{x}\right)\geq \text{0}\right\} _{i=\text{1}}^{I}\Longrightarrow f_{0}\left(\mathbf{x}\right)\geq \text{0}$
holds if and only if there exist $\lambda_{i}\geq \text{0},\forall i\in\mathcal{I}$
such that
\begin{equation}
\left[\begin{array}{cc}
\mathbf{A}_{0} & \mathbf{a}_{0}\\
\mathbf{a}_{0}^{H} & a_{0}
\end{array}\right]-\sum_{i=\text{1}}^{I}\lambda_{i}\left[\begin{array}{cc}
\mathbf{A}_{i} & \mathbf{a}_{i}\\
\mathbf{a}_{i}^{H} & a_{i}
\end{array}\right]\succeq\boldsymbol{0}.
\end{equation}
\label{lem2}
\end{lemma}

 We first address constraint $C$9, which involves the term $\left|\left(\mathbf{h}_{j,b}^{\mathrm{E}}\right)^{H}\boldsymbol{w}_{k,b}\right|^{\text{2}}$. To apply the $\mathcal{S}$-procedure, we need to express this term in quadratic form with respect to the channel uncertainty $\boldsymbol{\triangle}\boldsymbol{h}_{\mathrm{E},j}$.
Let $\mathbf{U}_{b}=\mathrm{diag}\left(\boldsymbol{u}_{b}\right)$ where $\boldsymbol{u}_{b}=\bar{\boldsymbol{\phi}}_{b}\odot\boldsymbol{\theta}$. Due to the coupled variables, we derive a lower bound on $\left|\left(\mathbf{h}_{j,b}^{\mathrm{E}}\right)^{H}\boldsymbol{w}_{k,b}\right|^{\text{2}}$ using first-order Taylor expansion at the $\tau$-th iteration of the AO, i.e., 
\begin{equation}
\left|\left(\mathbf{h}_{j,b}^{\mathrm{E}}\right)^{H}\boldsymbol{w}_{k,b}\right|^{\text{2}}\geq f_{\text{1}}+f_{\text{2}}-f_{\text{3}},
\end{equation}
where
\begin{align}
f_{\text{1}} &=\boldsymbol{h}_{\mathrm{E},j}^{H}\mathbf{U}_{b}^{\left(\tau\right)}\boldsymbol{G}\boldsymbol{w}_{k,b}^{\left(\tau\right)}\boldsymbol{w}_{k,b}^{H}\boldsymbol{G}^{H}\mathbf{U}_{b}^{H}\boldsymbol{h}_{\mathrm{E},j},\\
f_{\text{2}} &= f_{\text{1}}^{H},\\
f_{\text{3}} &=\boldsymbol{h}_{\mathrm{E},j}^{H}\mathbf{U}_{b}^{\left(\tau\right)}\boldsymbol{G}\boldsymbol{w}_{k,b}^{\left(\tau\right)}\left(\boldsymbol{w}_{k,b}^{\left(\tau\right)}\right)^{H}\boldsymbol{G}^{H}\left(\mathbf{U}_{b}^{\left(\tau\right)}\right)^{H}\boldsymbol{h}_{\mathrm{E},j}.
\end{align}

Then, substituting $\boldsymbol{h}_{\mathrm{E},j}=\bar{\boldsymbol{h}}_{\mathrm{E},j}+\boldsymbol{\triangle}\boldsymbol{h}_{\mathrm{E},j}$ and letting $\boldsymbol{x}_{j}=\boldsymbol{\triangle}\boldsymbol{h}_{\mathrm{E},j}$, we can express each term in the standard quadratic form as follows
\begin{equation}
f_{\ell}=\boldsymbol{x}_{j}^{H}\mathbf{A}_{\ell,k,b}\boldsymbol{x}_{j}+\text{2}\Re\left\{\mathbf{a}_{\ell,k,j,b}^{H}\boldsymbol{x}_{j}\right\}+a_{\ell,k,j,b}, \ell\in\{\text{1},\text{2},\text{3}\},
\end{equation}
where the matrices and vectors are defined as
\begin{align}
\mathbf{A}_{\text{1},k,b} &= \mathbf{U}_{b}^{\left(\tau\right)}\boldsymbol{G}\boldsymbol{w}_{k,b}\boldsymbol{w}_{k,b}^{H}\boldsymbol{G}^{H}\mathbf{U}_{b}^{H},\\
\mathbf{A}_{\text{2},k,b} &= \mathbf{A}_{\text{1},k,b}^{H},\\
\mathbf{A}_{\text{3},k,b} &= \mathbf{U}_{b}^{\left(\tau\right)}\boldsymbol{G}\boldsymbol{w}_{k,b}^{\left(\tau\right)}\left(\boldsymbol{w}_{k,b}^{\left(\tau\right)}\right)^{H}\boldsymbol{G}^{H}\left(\mathbf{U}_{b}^{\left(\tau\right)}\right)^{H},
\end{align}
and $\mathbf{a}_{\ell,k,j,b}=\mathbf{A}_{\ell,k,b}\bar{\boldsymbol{h}}_{\mathrm{E},j}$, $a_{\ell,k,j,b}=\bar{\boldsymbol{h}}_{\mathrm{E},j}^{H}\mathbf{A}_{\ell,k,b}\bar{\boldsymbol{h}}_{\mathrm{E},j}$ for $\ell\in\{\text{1},\text{2},\text{3}\}$. Combining these terms yields the quadratic lower bound
\begin{equation}
\left|\left(\mathbf{h}_{j,b}^{\mathrm{E}}\right)^{H}\boldsymbol{w}_{k,b}\right|^{\text{2}}\geq\boldsymbol{x}_{j}^{H}\mathbf{A}_{k,b}\boldsymbol{x}_{j}+\text{2}\Re\left\{ \mathbf{a}_{k,j,b}^{H}\boldsymbol{x}_{j}\right\} +a_{k,j,b},
\end{equation}
where $\mathbf{A}_{k,b}=\mathbf{A}_{\text{1},k,b}+\mathbf{A}_{\text{2},k,b}-\mathbf{A}_{\text{3},k,b}$, $\mathbf{a}_{k,j,b}=\mathbf{a}_{\text{1},k,j,b}+\mathbf{a}_{\text{2},k,j,b}-\mathbf{a}_{\text{3},k,j,b}$, and $a_{k,j,b}=\text{2}\Re\left\{ a_{\text{1},k,j,b}\right\} -a_{\text{3},k,j,b}$.

By applying the $\mathcal{S}$-procedure with the spherical uncertainty constraint $\|\boldsymbol{x}_{j}\|_{\text{2}}^{\text{2}}\leq\varepsilon_{j}^{\text{2}}$, constraint $C$9 is transformed to
 \begin{equation} \overline{C\text{9}}:\left[\begin{array}{cc}
\lambda_{k,j,b}^{\text{C9}}\mathbf{I}_{M}-\mathbf{A}_{k,b} & -\mathbf{a}_{k,j,b}\\
-\mathbf{a}_{k,j,b}^{H} & \bar{\lambda}_{k,j,b}^{\text{C9}}
\end{array}\right]\succeq\boldsymbol{0}_{M+\text{1}}, \forall k, \forall j, \forall b,
\end{equation}
 where $\lambda_{k,j,b}^{\text{C9}}\geq \text{0}$ and $\bar{\lambda}_{k,j,b}^{\text{C9}}=-a_{k,j,b}+\tilde{v}_{k,j,b}^{\mathrm{ub}}-\lambda_{k,j,b}^{\text{C9}}\varepsilon_{j}^{\text{2}}$ with $\tilde{v}_{k,j,b}^{\mathrm{ub}}=\left(\text{2}^{v_{k,b}^{\left(\tau\right)}}-\text{1}\right)\bar{v}_{k,j,b}+\bar{v}_{k,j,b}^{\left(\tau\right)}\text{2}^{v_{k,b}^{\left(\tau\right)}}\left(v_{k,b}-v_{k,b}^{\left(\tau\right)}\right)\ln\text{2}$ by adopting the SCA approach.

Following similar procedures, constraints $C$5$a$ and $C$10 can be transformed into LMI forms, which are given by
\begin{equation}
\overline{C\text{5}a}:\left[\begin{array}{cc} \mathbf{\tilde{A}}_{b}+\lambda_{j,b}^{\text{C5}}\mathbf{I}_{M} & \tilde{\mathbf{a}}_{j,b}\\ \mathbf{\tilde{a}}_{j,b}^{H} & \bar{\lambda}_{j,b}^{\text{C5}} \end{array}\right]\succeq\boldsymbol{0}_{M+\text{1}}, \forall j, \forall b,
\end{equation}
\begin{equation}
\overline{C\text{10}}:\left[\begin{array}{cc}
\mathbf{\tilde{A}}_{-k,b}+\lambda_{k,j,b}^{\text{C10}}\mathbf{I}_{M} & \mathbf{\tilde{a}}_{-k,j,b}\\
\mathbf{\tilde{a}}_{-k,j,b}^{H} & \bar{\lambda}_{-k,j,b}^{\text{C10}}
\end{array}\right]\succeq\boldsymbol{0}_{M+\text{1}}, \forall k, \forall j, \forall b,
\end{equation}
where $\lambda_{j,b}^{\text{C5}},\lambda_{k,j,b}^{\text{C10}}\geq \text{0}$, $\bar{\lambda}_{j,b}^{\text{C5}}=\tilde{a}_{j,b}-\Gamma_{j}^{\mathrm{E}}-\lambda_{j,b}^{\text{C5}}\varepsilon_{j}^{\text{2}}$, and $\bar{\lambda}_{-k,j,b}^{\text{C10}}=\tilde{a}_{-k,j,b}-\bar{v}_{k,j,b}+\sigma_{\mathrm{E},j}^{\text{2}}-\lambda_{k,j,b}^{\text{C10}}\varepsilon_{j}^{\text{2}}$. The aggregated matrices and vectors are constructed as
\begin{align}
\mathbf{\tilde{A}}_{b} &= \sum_{k=\text{1}}^{K}\chi_{k,b}\mathbf{A}_{k,b}+\mathbf{A}_{\mathrm{f},b},\\
\tilde{\mathbf{a}}_{j,b} &= \sum_{k=\text{1}}^{K}\chi_{k,b}\mathbf{a}_{k,j,b}+\mathbf{a}_{\mathrm{f},j,b},\\
\tilde{a}_{j,b} &= \sum_{k=\text{1}}^{K}\chi_{k,b}a_{k,j,b}+a_{\mathrm{f},j,b},
\end{align}
where $\mathbf{A}_{\mathrm{f},b}$, $\mathbf{a}_{\mathrm{f},j,b}$, and $a_{\mathrm{f},j,b}$ are obtained by replacing the beamforming vector $\boldsymbol{w}_{k,b}$ with the AN covariance matrix $\mathbf{f}_{b}$ in the expressions for $\mathbf{A}_{k,b}$, $\mathbf{a}_{k,j,b}$, and $a_{k,j,b}$, respectively. For constraint $C$10, the corresponding terms $\mathbf{\tilde{A}}_{-k,j,b}$, $\tilde{\mathbf{a}}_{-k,j,b}$, and $\tilde{a}_{-k,j,b}$ are obtained by excluding the $k$-th user's contribution from the aggregated expressions.

By replacing the semi-infinite constraints with their LMI equivalents, we obtain the following optimization problem
\begin{align}
P\text{3}:\, & \underset{\left\{ \boldsymbol{W}_{b},\mathbf{f}_{b}\right\} _{b=\text{1}}^{B},\boldsymbol{\phi},\boldsymbol{\theta},\boldsymbol{\chi},\mathbf{\Delta}}{\mathrm{max}}\enspace t\\
 & \mathrm{s.t.}\,C\text{1}-C\text{3},C\text{4}a,\overline{C\text{5}a},C\text{6}-C\text{8},\overline{C\text{9}},\overline{C\text{10}},\nonumber
\end{align}
where $\mathbf{\Delta}=\left\{t,\boldsymbol{z},\boldsymbol{\mu},\boldsymbol{v},\bar{\boldsymbol{v}},\boldsymbol{\lambda}\right\}$ represents all auxiliary variables with $\boldsymbol{\lambda}=\left\{ \lambda_{j,b}^{\text{C5}},\lambda_{k,j,b}^{\text{C9}},\lambda_{k,j,b}^{\text{C10}}|\forall k,j,b\right\}$.

Problem $P$3 remains non-convex due to the coupling among variables $\{\boldsymbol{W}_{b},\mathbf{f}_{b},\boldsymbol{\phi},\boldsymbol{\theta},\boldsymbol{\chi}\}$. This motivates the use of the AO method to decouple these variables. Specifically, AO alternately optimizes each variable block while fixing the others, thereby converting $P$3 into a sequence of convex subproblems. In each iteration, we update the transmit beamformer, AN, beam pattern assignment, and MRIS configurations in a cyclic manner until convergence.
\vspace{-3mm}
\subsection{Auxiliary Variables $\left\{\boldsymbol{z},\boldsymbol{\mu}\right\}$ Optimization}
 With other variables fixed, analytical solutions for $\left\{ z_{k,b}\right\} $ and $\left\{ \mu_{k,b}\right\} $ can be derived by setting the first-order derivatives of the cost function and $e_{k,b}$ to zero, respectively, which yield
 \begin{equation}
z_{k,b}=\text{1}+\mathrm{SINR}_{k,b}^{\mathrm{U}},
\end{equation}
\begin{equation}
\mu_{k,b}\!=\!\!\left(\left\Vert \!\left(\mathbf{h}_{k,b}^{\mathrm{U}}\right)^{H}\left(\underset{i\in\mathcal{K}}{\sum}\chi_{i,b}\boldsymbol{w}_{i,b}\!+\!\mathbf{f}_{b}\right)\!\right\Vert ^{\text{2}}\!\!\!+\!\sigma_{\mathrm{U},k}^{\text{2}}\right)^{-\text{1}}\!\!\!\!\!\left(\mathbf{h}_{k,b}^{\mathrm{U}}\right)^{H}\boldsymbol{w}_{k,b}.
\end{equation}
\vspace{-9mm}
\subsection{BS Beamformer $\left\{ \boldsymbol{w}_{k,b},\mathbf{f}_{b}\right\} $ Optimization}
In this subsection, to solve the subproblem of transmit beamforming for selected beam patterns $B_{\mathrm{s}}$, we fix the phase shift matrices of the MRIS in $P$3. Then, the problem of optimizing the BS beamformer and AN is
\begin{align}P\text{4}:\, & \underset{\left\{ \boldsymbol{W}_{b},\mathbf{f}_{b}\right\} _{b=\text{1}}^{B_{\mathrm{s}}},\mathbf{\Delta}}{\mathrm{max}}\enspace t\\
 & \mathrm{s.t.}\,C\text{1},\overline{C\text{5}a},\overline{C\text{9}},\overline{C\text{10}}\nonumber\\
 & \enspace\quad C\text{4}b:\sum_{b=\text{1}}^{B}\chi_{k,b}\left(\underset{i\in\mathcal{K}}{\sum}\chi_{i,b}\boldsymbol{w}_{i,b}^{H}\mathbf{S}_{k,b}\boldsymbol{w}_{i,b}+\mathbf{f}_{b}^{H}\mathbf{S}_{k,b}\mathbf{f}_{b}-\right.\nonumber\\
 & \enspace\quad\enspace\quad\left.\text{2}\Re\left\{ \mathbf{s}_{k,b}^{H}\boldsymbol{w}_{k,b}\right\} -\widetilde{z}_{k,b}\right)\leq-\Gamma_{k},\forall k,\nonumber\\
 & \enspace\quad C\text{8}a:\sum_{b=\text{1}}^{B}\chi_{k,b}\left(\underset{i\in\mathcal{K}}{\sum}\chi_{i,b}\boldsymbol{w}_{i,b}^{H}\mathbf{S}_{k,b}\boldsymbol{w}_{i,b}+\mathbf{f}_{b}^{H}\mathbf{S}_{k,b}\mathbf{f}_{b}-\right.\nonumber\\
 & \enspace\quad\enspace\quad\left.\text{2}\Re\left\{ \mathbf{s}_{k,b}^{H}\boldsymbol{w}_{k,b}\right\} -\widetilde{z}_{k,b}+v_{k,b}\right)\leq-t,\forall k, \nonumber
\end{align}
 where $\mathbf{S}_{k,b}\triangleq z_{k,b}\left|\mu_{k,b}\right|^{\text{2}}\mathbf{h}_{k,b}^{\mathrm{U}}\left(\mathbf{h}_{k,b}^{\mathrm{U}}\right)^{H}$ and  $\mathbf{s}_{k,b}=z_{k,b}\mu_{k,b}\mathbf{h}_{k,b}^{\mathrm{U}}$. 
 
 Since $P$4 is formulated as a convex optimization problem with linear objective, quadratic constraints, and LMI constraints, it can be efficiently solved using standard numerical solvers such as CVX.
\vspace{-3mm}
 \subsection{Beam Pattern Assignment $ \boldsymbol{\chi}$ Optimization}
By fixing all other variables, problem $P$3  reduces to the following beam pattern assignment subproblem
\begin{align}
P\text{5}:\, & \underset{\boldsymbol{\chi},\mathbf{\Delta}}{\mathrm{max}}\enspace t\\
 & \mathrm{s.t.}\,C\text{1}-C\text{3},C\text{4}a,\overline{C\text{5}a},C\text{8},\overline{C\text{9}},\overline{C\text{10}}.\nonumber  
\end{align}

Problem $P$5 presents several challenges due to the binary nature of scheduling variables and non-convex constraints. We address these systematically through relaxation and reformulation techniques.

 To facilitate efficient optimization, we first relax the binary scheduling variables $\chi_{k,b} \in \{\text{0},\text{1}\}$ to continuous values within $[\text{0},\text{1}]$, leading to the relaxed constraints $C\text{3}a:\text{0}\leq\chi_{k,b}\leq\text{1},\forall k,\forall b$ and $C\text{3}b:\sum_{b=\text{1}}^{B}\left(\chi_{k,b}-\chi_{k,b}^{\text{2}}\right)\leq \text{0},\forall k$. Constraint $C$3$b$ enforces binary-like behavior by penalizing fractional values, since $\chi_{k,b}-\chi_{k,b}^{\text{2}} = \chi_{k,b}(\text{1}-\chi_{k,b}) \geq \text{0}$ with equality only when $\chi_{k,b} \in \{\text{0},\text{1}\}$.

The bilinear term $\chi_{k,b}\sum_{i}\chi_{i,b}$ in $C\text{4}a$ and $C\text{8}$ is inherently non-convex, and this non-convexity remains even if the binary variables are relaxed to continuous. To deal with the mixed-integer-like structure, we employ the big-M \cite{r25} method, which converts conditional constraints into linear inequalities by introducing sufficiently large constants.
For constraint $C\text{4}a$, we introduce auxiliary variables $\varsigma_{k,b}$ and reformulate it as 
\begin{align}
C\text{4}c: &\enspace \varsigma_{k,b}\geq \text{0}, \forall k, b,\\
C\text{4}d: &\enspace c_{k,b}\!-\!\xi_{k,b}\!\left(\sum_{i\in\mathcal{K}}\chi_{i,b}\alpha_{k,i,b}\!+\!d_{k,b}\!\right)\!\geq\!\varsigma_{k,b}, \!\forall k,\! b,\\
C\text{4}e: &\enspace \varsigma_{k,b}\leq \mathrm{M}_{\text{1}}\chi_{k,b}, \forall k, b,\\
C\text{4}f: &\enspace \sum_{b=\text{1}}^{B}\varsigma_{k,b}\geq\Gamma_{k}, \forall k,
\end{align}
where $\mathrm{M}_{\text{1}}$ is a sufficiently large positive constant that ensures $\varsigma_{k,b} = \text{0}$ when $\chi_{k,b} = \text{0}$. We define  
\begin{equation}
    y_{k,b}=c_{k,b}-\xi_{k,b}\left(\underset{i\in\mathcal{K}}{\sum}\chi_{i,b}\alpha_{k,i,b}+d_{k,b}\right),
\end{equation} 
with $c_{k,b}=\text{2}z_{k,b}\Re\left\{ \mu_{k,b}^{\ast}\left(\mathbf{h}_{k,b}^{\mathrm{U}}\right)^{H}\boldsymbol{w}_{k,b}\right\} +\widetilde{z}_{k,b}$, 
$d_{k,b}=\left|\left(\mathbf{h}_{k,b}^{\mathrm{U}}\right)^{H}\mathbf{f}_{b}\right|^{\text{2}}$, 
$\alpha_{k,i,b}\!=\!\left|\!\left(\!\mathbf{h}_{k,b}^{\mathrm{U}}\!\right)^{H}\!\boldsymbol{w}_{i,b}\right|^{\text{2}}$, and $\xi_{k,b}\!=\!z_{k,b}\left|\mu_{k,b}\right|^{\text{2}}$.

Similarly, for constraint $C$8, we introduce auxiliary variables $r_{k,b}$ and obtain
\begin{align}
C\text{8}b: &\enspace r_{k,b}\geq \text{0}, \forall k, b,\\
C\text{8}c: &\enspace c_{k,b}\!\!-\!\xi_{k,b}\!\left(\!\sum_{i\in\mathcal{K}}\chi_{i,b}\alpha_{k,i,b}\!+\!\!d_{k,b}\!\!\right)\!\!-\!\!v_{k,b}\!\geq\! r_{k,b},\! \forall k,\! b,\\
C\text{8}d: &\enspace r_{k,b}\leq\mathrm{M}_{\text{2}}\chi_{k,b}, \forall k, b,\\
C\text{8}e: &\enspace t\leq\sum_{b=\text{1}}^{B}r_{k,b}, \forall k,
\end{align}
where $\mathrm{M}_{\text{2}}$ is another sufficiently large positive constant.

Since constraint $C$3$b$ remains non-convex due to the quadratic term $-\chi_{k,b}^{\text{2}}$, we apply the strong Lagrange duality approach by incorporating it into the objective function, resulting in the penalized optimization problem
\begin{align}
P\text{6}:\, & \underset{\boldsymbol{\chi},\mathbf{\Delta}}{\text{max}}\enspace t-\rho_{\text{1}}\sum_{b=\text{1}}^{B}\left(\chi_{k,b}-\chi_{k,b}^{\text{2}}\right)\\
 & \text{s.t.}\,C\text{1}, C\text{2}, C\text{3}a, C\text{4}c - C\text{4}f, \overline{C\text{5}a}, C\text{8}b - C\text{8}e,\overline{C\text{9}},\overline{C\text{10}},\nonumber 
\end{align}
 where $\rho_{\text{1}}\in[\text{0},\text{1}]$ is a sufficiently large penalty parameter that enforces binary-like behavior for $\chi_{k,b}$. The objective function in $P$6 remains non-convex due to the concave quadratic penalty terms $-\chi_{k,b}^\text{2}$. We linearize these terms using their first-order Taylor expansions around the point $\chi_{k,b}^{(\tau)}$. Therefore, a performance upper-bound of $P$6 can be acquired by solving
\begin{align}
P\text{7}\!:\, & \underset{\boldsymbol{\chi},\mathbf{\Delta}}{\text{max}}\enspace t\!-\!\rho_{\text{1}}\sum_{b=\text{1}}^{B}\left(\chi_{k,b}-\text{2}\chi_{k,b}^{\left(\tau\right)}\chi_{k,b}+\left(\chi_{k,b}^{\left(\tau\right)}\right)^{\text{2}}\right)  \\
 & \text{s.t.}\,C\text{1}, C\text{2}, C\text{3}a, C\text{4}c - C\text{4}f, \overline{C\text{5}a}, C\text{8}b - C\text{8}e,\overline{C\text{9}},\overline{C\text{10}},\nonumber 
\end{align}
which can be efficiently solved via CVX.
\vspace{-2mm}
\subsection{Phase Shifts $\boldsymbol{\phi}$ and $\boldsymbol{\theta}$ Optimization}
To optimize the phase-shift vectors $\boldsymbol{\phi}$ and $\boldsymbol{\theta}$ under nonconvex unit-modulus constraints while maintaining feasibility of the QoS/LMI-type constraints, we employ the PDD method \cite{r26}. Fixing all other variables and defining $\boldsymbol{\nu}_{\theta}=[e^{j\theta_{\text{1}}},\dots,e^{j\theta_{M}}]^{H}$, the corresponding optimization problem for the phase shift vector of S\text{1} can be formulated as
\begin{align}P\text{8}: & \underset{\boldsymbol{\nu}_{\theta},\mathbf{\Delta}}{\text{max}}\enspace t\\
 & \text{s.t.}\,\overline{C\text{5}a},C\text{6},\overline{C\text{9}},\overline{C\text{10}},\nonumber\\
 & \enspace\quad C\text{4}g:\boldsymbol{\nu}_{\theta}^{H}\mathbf{P}_{\text{1}}\boldsymbol{\nu}_{\theta}-\text{2}\Re\left\{ \mathbf{p}_{\text{1}}^{H}\boldsymbol{\nu}_{\theta}\right\} -\nonumber\\
 & \enspace\quad\enspace\quad\widetilde{z}_{k,b}+\Gamma_{k}\leq\text{0},\forall k,\forall b,\nonumber\\
 & \enspace\quad C\text{8}f:\boldsymbol{\nu}_{\theta}^{H}\mathbf{P}_{\text{1}}\boldsymbol{\nu}_{\theta}-\text{2}\Re\left\{ \mathbf{p}_{\text{1}}^{H}\boldsymbol{\nu}_{\theta}\right\} -\nonumber\\
 & \enspace\quad\enspace\quad\widetilde{z}_{k,b}+v_{k,b}\leq-t,\forall k,\forall b, \nonumber
\end{align}
where $\mathbf{p}_{\text{1},k,b}=z_{k,b}\mu_{k,b}^{\ast}\bar{\boldsymbol{\Phi}}_{b}\mathrm{diag}\left(\boldsymbol{h}_{\mathrm{U},k}^{H}\right)\boldsymbol{G}\boldsymbol{w}_{k,b}$, $\mathbf{P}_{\text{1},k,b}=z_{k,b}\left|\mu_{k,b}\right|^{\text{2}}\left(\bar{\boldsymbol{\Phi}}_{b}\mathrm{diag}\left(\boldsymbol{h}_{\mathrm{U},k}^{H}\right)\boldsymbol{G}\mathbf{x}_{b}\mathbf{x}_{b}^{H}\boldsymbol{G}^{H}\mathrm{diag}\left(\boldsymbol{h}_{\mathrm{U},k}\right)\bar{\boldsymbol{\Phi}}_{b}^{H}\right)$, and $\mathbf{x}_{b}=\left\{ \chi_{k,b}\boldsymbol{w}_{k,b},\mathbf{f}_{b}|\forall k\right\}$.

Conventional semidefinite relaxation methods may fail to guarantee feasibility since QoS constraints could be violated if the obtained solution does not exhibit a rank-one structure. The PDD method decomposes the unit-modulus constraints by introducing auxiliary variables and an augmented Lagrangian formulation, providing superior numerical stability compared to direct penalty approaches.
To tackle the difficult nonlinear equality constraint $C\text{6}$, we introduce
auxiliary variables $\breve{\boldsymbol{\nu}}_{\theta}$ and rewrite the
problem $P$8 as follows
\begin{align}P\text{9}: & \underset{\boldsymbol{\nu}_{\theta},\breve{\boldsymbol{\nu}}_{\theta},\mathbf{\Delta}}{\text{max}}\enspace t\\
 & \text{s.t.}\,C\text{4}g,\overline{C\text{5}a},C\text{8}f,\overline{C\text{9}},\overline{C\text{10}},\nonumber\\
 & \enspace\quad C\text{11}:\boldsymbol{\nu}_{\theta}=\breve{\boldsymbol{\nu}}_{\theta},\nonumber\\
 & \enspace\quad C\text{12}:\left|\breve{\boldsymbol{\nu}}_{\theta,m}\right|=\text{1},\forall m,\nonumber\\
 & \enspace\quad C\text{13}:\left|\boldsymbol{\nu}_{\theta,m}\right|\leq1,\forall m.\nonumber
\end{align}

The equality constraint $C$11 is then incorporated into the objective function using the augmented Lagrangian method
\begin{align}P\text{10}: & \underset{\boldsymbol{\nu}_{\theta},\breve{\boldsymbol{\nu}}_{\theta},\mathbf{\Delta}}{\text{max}}\enspace t\!-\!\!\left(\!\frac{\text{1}}{\text{2}\rho_{\text{2}}}\left\Vert \boldsymbol{\nu}_{\theta}\!-\!\breve{\boldsymbol{\nu}}_{\theta}\right\Vert _{\text{2}}^{\text{2}}\!+\!\Re\left\{ \boldsymbol{\lambda}_{\theta}^{H}\left(\boldsymbol{\nu}_{\theta}\!-\!\breve{\boldsymbol{\nu}}_{\theta}\right)\!\right\} \right)\\
 & \text{s.t.}\,C\text{4}g,\overline{C\text{5}a},C\text{8}f,\overline{C\text{9}},\overline{C\text{10}},C\text{12},C\text{13}, \nonumber
\end{align}
where $\rho_{\text{2}}$ and $\boldsymbol{\lambda}_{\theta}$ are penalty parameter and dual variable, respectively.

The PDD algorithm alternates between two subproblems. When $\breve{\boldsymbol{\nu}}_{\theta}$ is fixed, we solve the following inner loop convex subproblem for $\boldsymbol{\nu}_{\theta}$, i.e., 
\begin{align}P\text{11}: & \underset{\boldsymbol{\nu}_{\theta},\mathbf{\Delta}}{\text{max}}\enspace t\!-\!\!\left(\!\frac{\text{1}}{\text{2}\rho_{\text{2}}}\left\Vert \boldsymbol{\nu}_{\theta}\!-\!\breve{\boldsymbol{\nu}}_{\theta}\right\Vert _{\text{2}}^{\text{2}}\!+\!\Re\left\{ \boldsymbol{\lambda}_{\theta}^{H}\left(\boldsymbol{\nu}_{\theta}\!-\!\breve{\boldsymbol{\nu}}_{\theta}\right)\!\right\} \right)\\
 & \text{s.t.}\,C\text{4}g,\overline{C\text{5}a},C\text{8}f,\overline{C\text{9}},\overline{C\text{10}},C\text{13}. \nonumber
\end{align}

 When $\boldsymbol{\nu}_{\theta},\mathbf{\Delta}$  are fixed, the auxiliary variables are updated by solving
\begin{align}
P\text{12}: & \underset{\left|\breve{\boldsymbol{\nu}}_{\theta,m}\right|=\text{1}}{\text{min}}\enspace\Re\left\{ \left(\rho_{\text{2}}^{-\text{1}}\boldsymbol{\nu}_{\theta}+\boldsymbol{\lambda}_{\theta}\right)^{H}\breve{\boldsymbol{\nu}}_{\theta}\right\}  
\end{align}
which admits the closed-form solution $\breve{\boldsymbol{\nu}}_{\theta}^{\star}=\exp\left(j\times\angle\left(\rho_{\text{2}}^{-\text{1}}\boldsymbol{\nu}_{\theta}+\boldsymbol{\lambda}_{\theta}\right)\right)$ by aligning the phases of $\breve{\boldsymbol{\nu}}_{\theta}$ with the argument of the gradient term.

The outer loop manages updates to the dual variable and penalty parameter based on constraint violation. When the equality constraint $\boldsymbol{\nu}_{\theta}=\breve{\boldsymbol{\nu}}_{\theta}$ is approximately satisfied, the dual variable is updated as $\boldsymbol{\lambda}_{\theta}\leftarrow\boldsymbol{\lambda}_{\theta}+\rho_{\text{2}}^{-\text{1}}\left(\boldsymbol{\nu}_{\theta}-\breve{\boldsymbol{\nu}}_{\theta}\right)$. Otherwise, the penalty parameter is increased as $\rho_{\text{2}}^{-\text{1}}\leftarrow\varpi_{\text{1}}^{-\text{1}}\times\rho_{\text{2}}^{-\text{1}}$ with $\varpi_{\text{1}}\in[\text{0.8},\text{0.9}]$ to enforce better constraint satisfaction.

For the S2 phase shift optimization, we employ an identical PDD framework to tackle the difficult nonlinear
equality constraint $C$7. We introduce $\boldsymbol{\nu}_{\phi}=[e^{j\phi_{\text{1}}},\dots,e^{j\phi_{N}}]^{H}$ and define the beam-dependent equivalent vector $\bar{\boldsymbol{\nu}}_{\phi,b}=\boldsymbol{E}_{b}\boldsymbol{\nu}_{\phi}+\boldsymbol{e}_{b}$, which maps the S2 elements to their corresponding positions on S1 for each beam pattern $b$. The optimization problem of the inner loop becomes
\begin{align}P\text{13}: & \underset{\boldsymbol{\nu}_{\phi},\mathbf{\Delta}}{\text{max}}\quad t\!-\!\!\left(\!\frac{\text{1}}{\text{2}\rho_{\text{3}}}\left\Vert \boldsymbol{\nu}_{\phi}\!-\!\breve{\boldsymbol{\nu}}_{\phi}\right\Vert _{\text{2}}^{\text{2}}\!+\!\Re\left\{ \boldsymbol{\lambda}_{\phi}^{H}\left(\boldsymbol{\nu}_{\phi}\!-\!\breve{\boldsymbol{\nu}}_{\phi}\right)\!\right\} \right)\\
 & \text{s.t.}\,\overline{C\text{5}a},\overline{C\text{9}},\overline{C\text{10}},\nonumber\\
 & \enspace\quad C\text{4}h:\bar{\boldsymbol{\nu}}_{\phi,b}^{H}\mathbf{P}_{\text{2},k,b}\bar{\boldsymbol{\nu}}_{\phi,b}-\text{2}\Re\left\{ \mathbf{p}_{\text{2},k,b}^{H}\bar{\boldsymbol{\nu}}_{\phi,b}\right\} -\nonumber\\
 & \enspace\quad\enspace\quad\widetilde{z}_{k,b}+\Gamma_{k}\leq\text{0},\forall k,\forall b,\nonumber\\
 & \enspace\quad C\text{8}g:\bar{\boldsymbol{\nu}}_{\phi,b}^{H}\mathbf{P}_{\text{2},k,b}\bar{\boldsymbol{\nu}}_{\phi,b}-\text{2}\Re\left\{ \mathbf{p}_{\text{2},k,b}^{H}\bar{\boldsymbol{\nu}}_{\phi,b}\right\} - \nonumber\\
 & \enspace\quad\enspace\quad\widetilde{z}_{k,b}+v_{k,b}\leq-t,\forall k,\forall b, \nonumber\\
 & \enspace\quad C\text{14}:\left|\boldsymbol{\nu}_{\phi,n}\right|\leq1,\forall n.\nonumber
\end{align}
where $\mathbf{p}_{\text{2},k,b}=z_{k,b}\mu_{k,b}^{\ast}\boldsymbol{\Theta}\mathrm{diag}\left(\boldsymbol{h}_{\mathrm{U},k}^{H}\right)\boldsymbol{G}\boldsymbol{w}_{k,b}$, $\mathbf{P}_{\text{2},k,b}=z_{k,b}\left|\mu_{k,b}\right|^{\text{2}}\left(\boldsymbol{\Theta}\mathrm{diag}\left(\boldsymbol{h}_{\mathrm{U},k}^{H}\right)\boldsymbol{G}\mathbf{x}_{b}\mathbf{x}_{b}^{H}\boldsymbol{G}^{H}\mathrm{diag}\left(\boldsymbol{h}_{\mathrm{U},k}\right)\boldsymbol{\Theta}^{H}\right)$, and the dual variable $\boldsymbol{\lambda}_{\phi}$ and penalty parameter $\rho_{\text{3}}$ are updated in the outer loop following the above rules.

Although the phase shifts optimization of S2 involves beam-dependent coupling through the linear transformation $\boldsymbol{E}_{b}$, this deterministic mapping does not fundamentally alter the PDD algorithm structure. The matrices $\boldsymbol{E}_{b}$ and offset vectors $\boldsymbol{e}_{b}$ are predetermined constants that define the geometric relationship between movable surface positions and beam patterns. Consequently, the quadratic constraints in $P$13 can be efficiently handled by iterating over all selected beam indices $b \in \mathcal{B}$ while maintaining the exact auxiliary variable decomposition and augmented Lagrangian formulation as in the S1 case. The unit-modulus constraints on $\boldsymbol{\nu}_{\phi}$ are enforced through identical dual variable updates and penalty parameter adjustments, ensuring algorithmic consistency. The computational complexity remains polynomial in the number of S2 elements $N$, with the beam-dependent structure introducing only linear overhead proportional to the number of beam patterns $B$.

\begin{algorithm}[t]
\caption{PDD Method to Optimize Phase Shifts $\boldsymbol{\theta}$}
\label{alg:pdd}
\begin{algorithmic}[1]
\State Initial $\boldsymbol{\nu}_{\theta}^{(\text{0})}$, $\breve{\boldsymbol{\nu}}_{\theta}^{(\text{0})}$,  $\boldsymbol{\lambda}_{\theta}^{(\text{0})}$,  $\rho_{2}^{(\text{0})}$, $\boldsymbol{v}^{(\text{0})}$, $\bar{\boldsymbol{v}}^{(\text{0})}$; convergence tolerances $\varrho_{\mathrm{out}}$, $\varrho_{\mathrm{inn}}$; penalty-update threshold $\tilde{\varrho}^{(\text{0})}$; $t \gets \text{0}$;
\Repeat 
    \State set $\boldsymbol{\nu}_{\theta}^{(t-1,\text{0})} \gets \boldsymbol{\nu}_{\theta}^{(t-1)}$,  $\breve{\boldsymbol{\nu}}_{\theta}^{(t-1,\text{0})} \gets \breve{\boldsymbol{\nu}}_{\theta}^{(t-1)}$,  $\iota \gets \text{0}$;
    \Repeat  
        \State update $\boldsymbol{\nu}_{\theta}^{(t-1,\iota+1)}$ by solving problem $P$11;
        \State update $\breve{\boldsymbol{\nu}}_{\theta}^{(t-1,\iota+1)}$ by the closed-form projection;
        \State $\iota \gets \iota + 1$;
    \Until{$\left\|\boldsymbol{\nu}_{\theta}^{(t-1,\iota)}-\boldsymbol{\nu}_{\theta}^{(t-1,\iota-1)}\right\|_{2}\le \varrho_{\mathrm{inn}}$}
    \State set $\boldsymbol{\nu}_{\theta}^{(t)} \gets \boldsymbol{\nu}_{\theta}^{(t-1,\infty)}$, $\breve{\boldsymbol{\nu}}_{\theta}^{(t)} \gets \breve{\boldsymbol{\nu}}_{\theta}^{(t-1,\infty)}$;
    \If{$\left\|\boldsymbol{\nu}_{\theta}^{(t)}-\breve{\boldsymbol{\nu}}_{\theta}^{(t)}\right\|_{\infty}\le \tilde{\varrho}^{(t)}$}
        \State $\boldsymbol{\lambda}_{\theta}^{\left(t+1\right)} \gets \boldsymbol{\lambda}_{\theta}^{\left(t\right)}+\frac{1}{\rho_{\text{2}}^{\left(t\right)}}\left(\boldsymbol{\nu}_{\theta}^{\left(t\right)}-\breve{\boldsymbol{\nu}}_{\theta}^{\left(t\right)}\right)$, $\rho_{\text{2}}^{\left(t+1\right)} \gets \rho_{\text{2}}^{\left(t\right)}$;
    \Else
        \State $\boldsymbol{\lambda}_{\theta}^{\left(t+1\right)} \gets \boldsymbol{\lambda}_{\theta}^{\left(t\right)}$, $\rho_{\text{2}}^{\left(t+1\right)}\gets\varpi_{\text{1}}\rho_{\text{2}}^{\left(t\right)}$;
    \EndIf
    \State $t \gets t+\text{1}$;
\Until{$\left\|\boldsymbol{\nu}_{\theta}^{(t)}-\breve{\boldsymbol{\nu}}_{\theta}^{(t)}\right\|_{2}\le \varrho_{\mathrm{out}}$}
\end{algorithmic}
\vspace{-1mm}
\end{algorithm}
\vspace{-2mm}
\subsection{Complexity and Convergence Analysis}
Based on the above description, the proposed PDD framework is summarized in Algorithm~\ref{alg:pdd}, and the overall AO procedure is presented in Algorithm \ref{alg:AO}. The variables $\left(\boldsymbol{W}_{b},\mathbf{f}_{b},\boldsymbol{\phi},\boldsymbol{\theta},\boldsymbol{\chi}\right)$ are alternately optimized until convergence. All subproblems are convex and involve linear, second-order cone (SOC), and LMI constraints, which can be solved by the interior point method \cite{r23}. Hence, the overall complexity of the AO algorithm can be evaluated as follows. The problem $P$4 contains SOC and the LMIs constraints with block size $\left(M+1\right)\times\left(M+1\right)$ and $\left(\text{2}K_{b}+1\right)J$ blocks, where $K_{b}$ is the number of users selected the beam pattern $b$. Its complexity is given by $S_{\mathbf{Wf}}\left(b\right)=\mathcal{O}\left(\left(\text{2}K_{b}+1\right)J\left(M+1\right)^{\text{3}}+\left(\text{2}L\left(K_{b}+1\right)+\left(\text{2}K_{b}+1\right)J\right)^{\text{3}}\right)$.
Problem $P$7 includes $\left(\text{2}K+1\right)JB$ LMI blocks and $\mathcal{O}\left(KB\right)$ variables, resulting in the complexity $S_{\mathbf{\mathbf{\chi}}}\!=\!\mathcal{O}\left(\left(\text{2}K+1\right)JB\left(M+1\right)^{\text{3}}+\left(KB\right)^{\text{3}}\right)$. The subproblems of S1 and S2 have the same order of complexity  $S_{\mathbf{\mathbf{\theta}}}=S_{\mathbf{\mathbf{\phi}}}=\mathcal{O}\left(\left(\text{2}K_{b}+1\right)J\left(M+1\right)^{\text{3}}\right)$, where the additional polynomial terms in $M$ and $N$ are negligible compared with the LMI cost. Accordingly, the each iteration complexity of the AO framework can be expressed as $S_{\mathrm{AO}}=\sum_{b=\textrm{1}}^{B_{\mathrm{s}}}S_{\mathbf{Wf}}\left(b\right)+S_{\mathbf{\mathbf{\chi}}}+I_{\textrm{PDD}}^{\left(\theta\right)}S_{\mathbf{\mathbf{\theta}}}+I_{\textrm{PDD}}^{\left(\phi\right)}S_{\mathbf{\mathbf{\phi}}}$, where $I_{\textrm{PDD}}$ is the iterations of  PDD algorithms, and the total complexity is $S=\mathcal{O}\left(I_{\textrm{AO}}S_{\mathrm{AO}}\right)$ with iterations of the AO $I_{\textrm{AO}}$. Each convex subproblem ensures a non-decreasing objective value, and is bounded by the finite system power, implying convergence of the proposed algorithm to a stable point.

\section{Simulation Results}
In this section, simulations are performed to evaluate the proposed scheme. The number of transmit and sensing antennas at the BS is $L=8$. The BS antennas are arranged in a vertical uniform linear array orthogonal to the ground plane, ensuring directional control in the elevation domain. The transmit power of the BS is $P_{\mathrm{max}}=\text{20}$ dBm. The BS and MRIS are located at $\left(\text{0},\text{0},\text{10}\right)$ m and $\left(\text{0},\text{10},\text{15}\right)$ m, respectively. For the spatial distribution, we configure the minimum distances of the inter-user, inter-Eve, and user-Eve as $d_{\mathrm{UU}} = 5$ m,  $d_{\mathrm{EE}} = 10$ m, and $d_{\mathrm{UE}} = 8$ m within the range $x \in [-\text{20},\text{20}]$ m, $y \in [\text{20},\text{70}]$ m, and $z \in [\text{0},\text{10}]$ m, respectively. The noise power of the user and Eve are $\sigma_{\mathrm{U}}^{\text{2}}=\sigma_{\mathrm{E}}^{\text{2}}=-\text{80}$ dBm and the path loss is $\beta_0=-\text{30}$ dBm. The CSI uncertainty of the Eve's channel is $\epsilon_{\mathrm{E},j,1}=\epsilon_{\mathrm{E},j,\text{2}}=\cdots=\epsilon_{\mathrm{E},j,m}=\text{0.1}\sqrt{\kappa_{\mathrm{RE},j}}$ with $\kappa_{\mathrm{RE},j}=\kappa_{\mathrm{RU},k}=\kappa_{\mathrm{BR}}=\text{5 dB},\forall k,\forall j$ denoting the Rician factors.  To evaluate the complex uncertainty in the target angle, distance, and NLoS parameters, we
consider the error bounds are $D_{\mathrm{RE},j} = $ 1 m, $\varTheta_{\mathrm{RE},j}=\text{1}^\circ$, and $\varPsi_{\mathrm{RE},j}=\text{1}^\circ$, respectively.

\begin{algorithm}[t]
\caption{Proposed AO Framework for Problem $P$1}
\label{alg:AO}
\begin{algorithmic}[1]
\State Initial $\boldsymbol{W}_{b}^{\left(\text{0}\right)}$, $\mathbf{f}_{b}^{\left(\text{0}\right)}$,
$\boldsymbol{\phi}^{\left(\text{0}\right)}$, $\boldsymbol{\theta}^{\left(\text{0}\right)}$,
$\boldsymbol{\chi}^{\left(\text{0}\right)}$, $\boldsymbol{v}^{\left(\text{0}\right)}$,
$\overline{\boldsymbol{v}}^{\left(\text{0}\right)}$; set the convergence tolerance
$\varrho_{\mathrm{AO}}$, maximum number of iterations $\tau_{\max}$; $\tau\gets\text{0}$;
\Repeat
    \If{$\tau\leq\tau_{\max}$}
        \State update $\left\{ \boldsymbol{W}_{b}^{\left(\tau\right)},\mathbf{f}_{b}^{\left(\tau\right)}\right\}_{b=\text{1}}^{B_{\mathrm{s}}}$ by solving $P$4;
        \State update $\boldsymbol{\theta}^{\left(\tau\right)}$ and $\boldsymbol{\phi}^{\left(\tau\right)}$ by using Algorithm \ref{alg:pdd};
        \State update $\boldsymbol{\chi}^{\left(\tau\right)}$ by solving $P$7;
        \State $\tau\gets\tau+\text{1}$;
    \EndIf
\Until{$\left|\left(t^{\left(\tau\right)}-t^{\left(\tau-\text{1}\right)}\right)/t^{\left(\tau\right)}\right|\leq\varrho_{\mathrm{AO}}$}
\end{algorithmic}
\end{algorithm}

\begin{figure}[!t]
\centering
\includegraphics[width=3.4in,trim=0 0 0 0, clip]{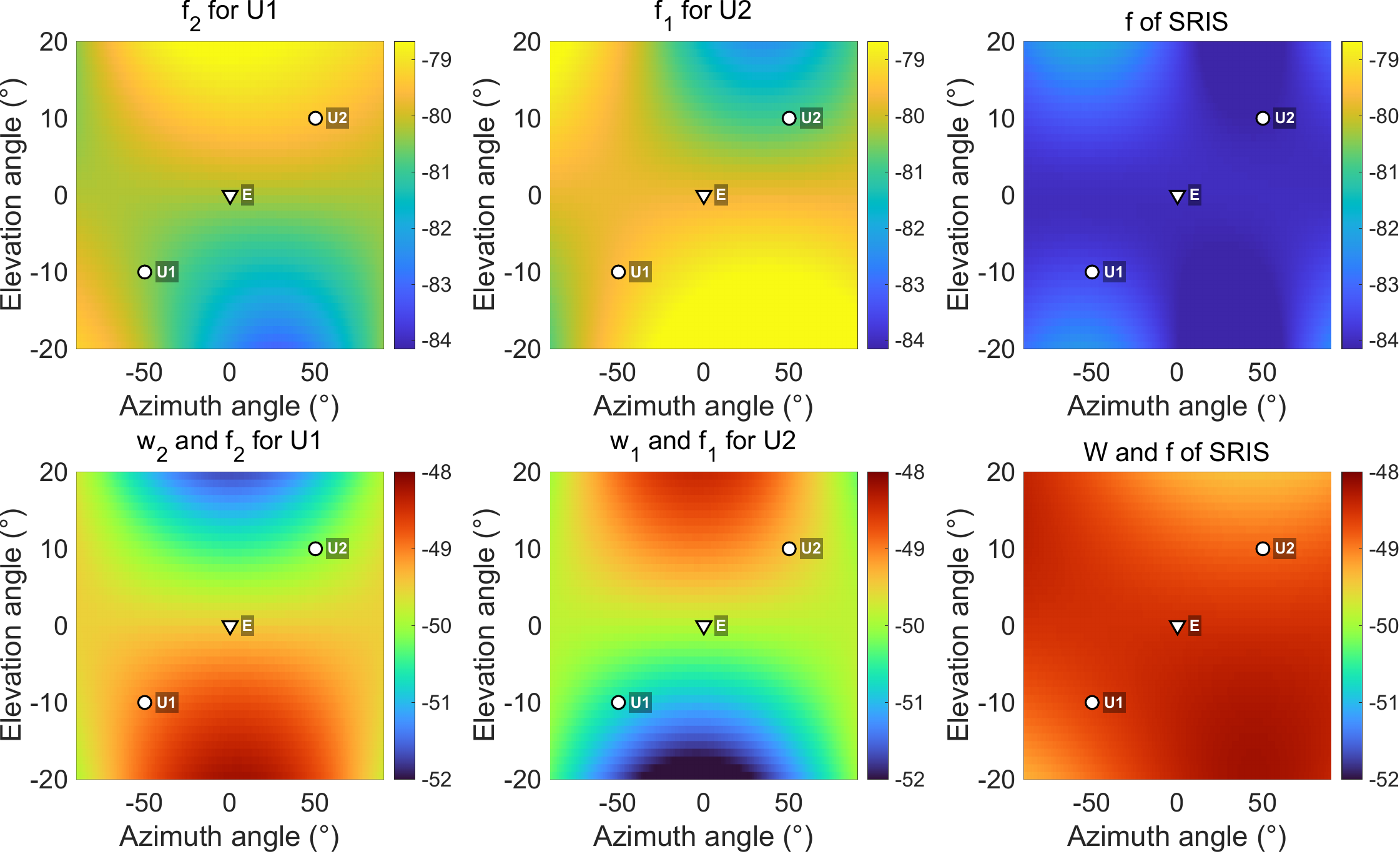}
\caption{Beam patterns achieved by the MRIS and a single-layer SRIS with respect to azimuth and elevation angles ($M = \text{2}\times\text{2}$, $N = \text{1}\times\text{2}$).}
\label{fig4}
\vspace{-4mm}
\end{figure}

Fig. \ref{fig4} illustrates the beam pattern comparison between the MRIS and a conventional single-layer SRIS. The MRIS is configured with $M = \text{2}\times\text{2}$ elements for S1 and $N = \text{1}\times\text{2}$ elements for S2, which can generate two distinct beam patterns. The system involves two users located at $\left(\text{60 m},-\text{50}^\circ, -\text{10}^\circ \right)$ and  $\left(\text{60 m}, \text{50}^\circ, \text{10}^\circ \right)$, and one Eve located at $\left(\text{60 m},\text{0}^\circ, \text{0}^\circ \right)$, as observed from the MRIS. The minimum secrecy rates achieved by the MRIS-and SRIS-assisted schemes are 1.51 bps/Hz and 0.44 bps/Hz, respectively, confirming the substantial improvement in security performance provided by the MRIS. Fig. \ref{fig4} presents the beam-gain distributions as a function of azimuth and elevation angles. The top row corresponds to the AN signal, while the bottom row shows the total gain combining the communication and AN components. In the MRIS case, user 1 (U1) is served by beam 2, and user 2 (U2) is served by beam 1, whereas in the SRIS, a single fixed beam simultaneously covers both users. One can find that the MRIS forms two high-gain beams precisely aligned with the users. At the same time, the AN energy is steered toward the Eve to create deliberate interference and enable target sensing, thereby effectively suppressing information leakage. In contrast, the SRIS can form only one fixed beam with limited angular adaptability, leading to dispersed energy and degraded secrecy performance. Overall, the MRIS can adaptively reshape its beam gain distribution according to the spatial positions of users and Eves. Although the MRIS employs static phase configurations, the S2 provides an additional geometric DoF, enabling mechanical beam steering and beam scheduling. This capability allows the MRIS to maintain reliable communication for users while simultaneously enhancing sensing and interference toward Eves.

\begin{figure}[!t]
\centering
\includegraphics[width=2.8in,trim=0 0 0 0, clip]{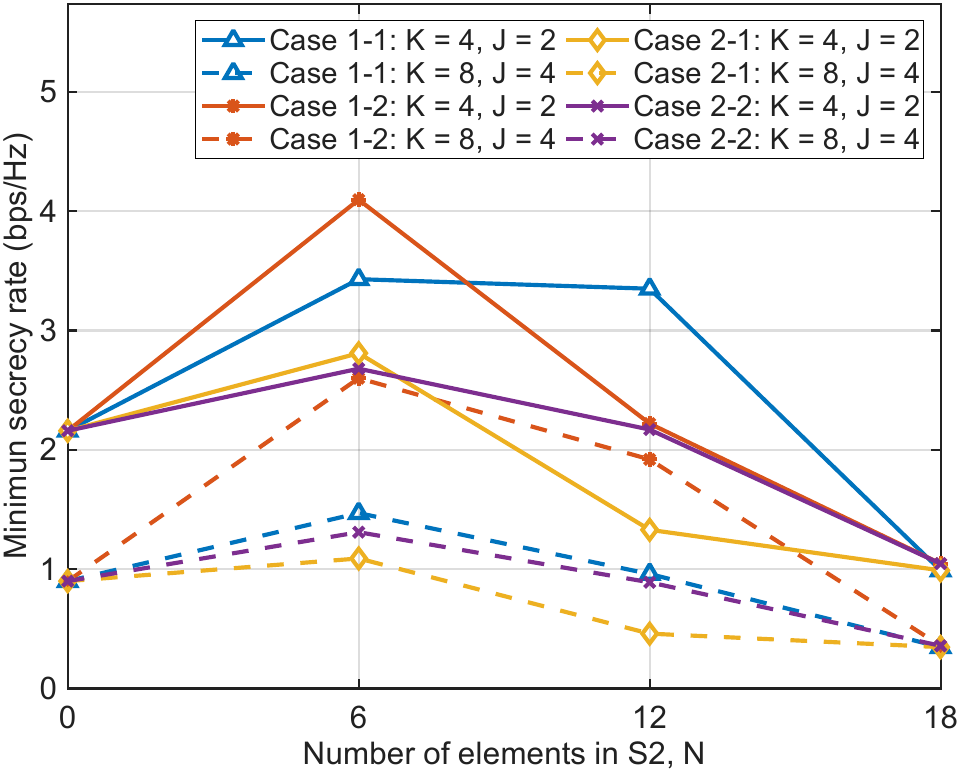}
\caption{Minimum secrecy rate performance under
different strategies of element allocation between S1 and S2 ($M+N$ = 36).}
\label{fig5}
\vspace{-3mm}
\end{figure} 

Fig. \ref{fig5} illustrates the system minimum secrecy rate performance under different element-allocation strategies between S1 and S2, with the total number of refracting elements fixed at $M+N=\text{36}$. Four allocation strategies are considered. Case 1-1 reallocates S1’s rows to S2 in a two-dimensional (2D) layout, where S1 decreases from $\text{6}\times\text{6}$ to $\text{3}\times\text{6}$ while S2 increases from $\text{0}\times\text{3}$ to $\text{3}\times\text{6}$. Case 1-2 follows a similar procedure along the column dimension. Case 2-1 performs one-dimensional (1D) reallocation, with S1 reducing from $\text{6}\times\text{6}$ to $\text{3}\times\text{6}$ and S2 expanding from $\text{0}\times\text{6}$ to $\text{3}\times\text{6}$. Case 2-2 conducts the same operation along the column direction. For each allocation mode, two system configurations are examined, namely $\left(K, J\right)=\left(\text{4}, \text{2}\right)$ and $\left(K, J\right)=\left(\text{8}, \text{4}\right)$. As shown in Fig. \ref{fig5}, the secrecy rate in all cases exhibits a unimodal trend, initially increasing and then decreasing as more elements are transferred from S1 to S2, with the peak occurring around $N=$6. When only a small number of elements are assigned to S2, the secrecy rate improves noticeably because S1 maintains a large effective aperture for coherent gain. At the same time, S2 provides additional phase-adjustable degrees of freedom. This configuration allows the MRIS to flexibly reshape its reflected beams, reinforcing legitimate links and suppressing eavesdropping. As the number of elements on S2 continues to grow, the aperture of S1 shrinks, its focusing capability weakens, and the overall energy distribution becomes less concentrated, leading to reduced secrecy performance. This loss of focusing capability can lead to performance even worse than that of a conventional SRIS with the same total number of elements ($N=0$). Among the four cases, the 2D reallocations consistently outperform the 1D ones. In particular, Case 1-2 (column-direction allocation) achieves the highest peak secrecy rate, followed by Case 1-1 (row-direction allocation). This difference is mainly due to the system geometry and the vertical configuration of the BS antenna array, which means that the BS primarily controls elevation-domain beamforming, while column-direction allocation enhances azimuth-domain controllability. The resulting spatial complementarity enables finer three-dimensional beam shaping and more effective interference suppression. As the number of users and eavesdroppers increases—from $\left(K, J\right)=\left(\text{4}, \text{2}\right)$ to  $\left(K, J\right)=\left(\text{8}, \text{4}\right)$, the overall secrecy rates decrease due to increased resource competition, but the relative trends remain consistent. In summary, allocating a small number of elements to S2
 preserves the aperture gain of S1 while introducing spatial reconfigurability, achieving a balanced design that offers superior three-dimensional beam focusing and secrecy performance.

\begin{figure}[!t]
\centering
\includegraphics[width=2.8in,trim=0 0 0 0, clip]{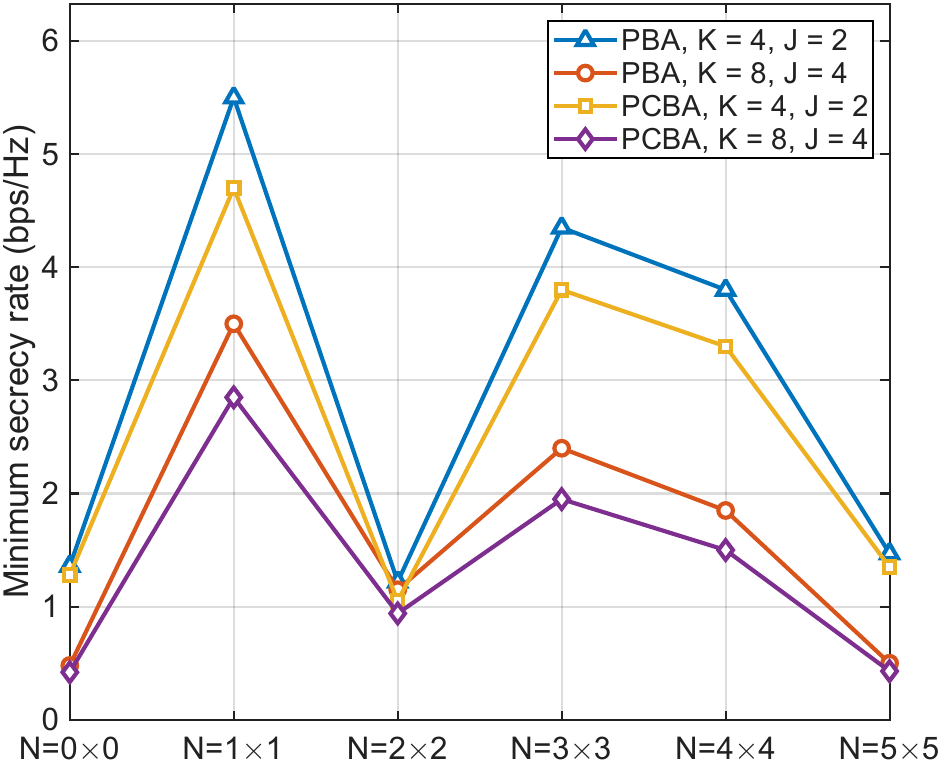}
\caption{Minimum secrecy rate performance over the single-layer SRIS with increasing $N_\mathrm{r}\times N_\mathrm{c}$ ($M=\text{5}\times \text{5}$).}
\label{fig6}
\vspace{-3mm}
\end{figure}

\begin{figure}[!t]
\centering
\includegraphics[width=2.8in,trim=0 0 0 0, clip]{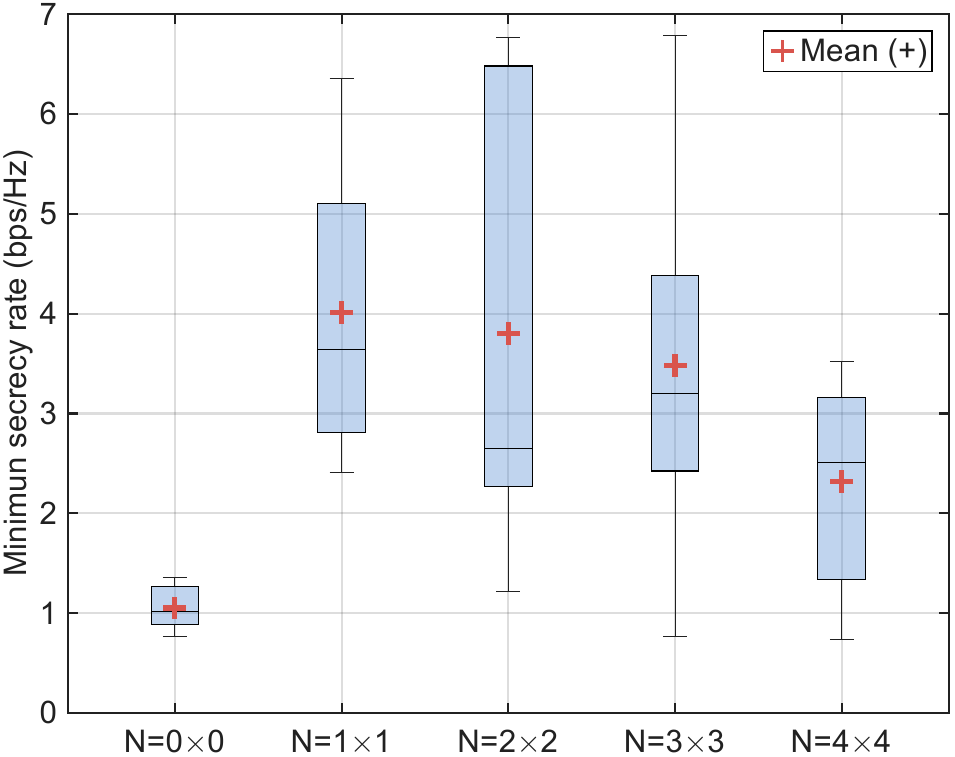}
\caption{Minimum secrecy rate distribution of the proposed PDD-based algorithm under 12 random realizations of user and Eve locations ($M=\text{5}\times\text{5}$, $K=\text{4}$, $J=\text{2}$)}
\label{fig7}
\vspace{-3mm}
\end{figure}

Fig. \ref{fig6} compares the minimum secrecy rate performance of the proposed PDD-based algorithm (PBA) and the benchmark penalized convex-concave programming-based algorithm (PCBA)\cite{r31} as the number of S2 elements increases from $N=\text{0}\times\text{0}$ to $N=\text{5}\times\text{5}$, with the S1 size fixed at $M=\text{5}\times\text{5}$.
The two algorithms exhibit a similar performance trend with respect to $N$, and PBA consistently achieves higher secrecy rates than PCBA across all configurations. This confirms the effectiveness of the proposed optimization framework. The minimum secrecy rate reaches its peak at $N=\text{1}\times\text{1}$, which proves that the secrecy rate improves significantly when only a small number of elements are assigned to S2. However, it decreases as more elements are added, similar to Fig. 5. A noticeable performance drop appears at $N=\text{2}\times\text{2}$, which deviates from the overall monotonic behavior. To investigate this anomaly, Fig. \ref{fig7} presents the statistical distribution of the minimum secrecy rate across 12 random realizations of the user and Eve locations in the 3D space. In Fig. \ref{fig7},  each box represents the interquartile range (25th to 75th percentiles), the central line denotes the median, whiskers indicate the extrema excluding outliers, and the "+" marker corresponds to the mean. The plot confirms that both the median and the mean values reach their maximum at $N=\text{1}\times\text{1}$ and gradually decline as $N$ increases, confirming that the best performance occurs under a small-scale S2 configuration. All MRIS cases ($N>\text{0}$) outperform the single-layer SRIS baseline ($N=\text{0}\times\text{0}$) in both median and mean values. Moreover, the $N=\text{2}\times\text{2}$ configuration shows a notably wider variance and interquartile range, indicating strong sensitivity to the geometric arrangement of the system nodes. This variability explains the fluctuation observed in Fig. \ref{fig6} and reveals that the MRIS secrecy performance is inherently geometry-dependent under certain configurations.

\begin{figure}[!t]
\centering
\includegraphics[width=2.8in,trim=0 0 0 0, clip]{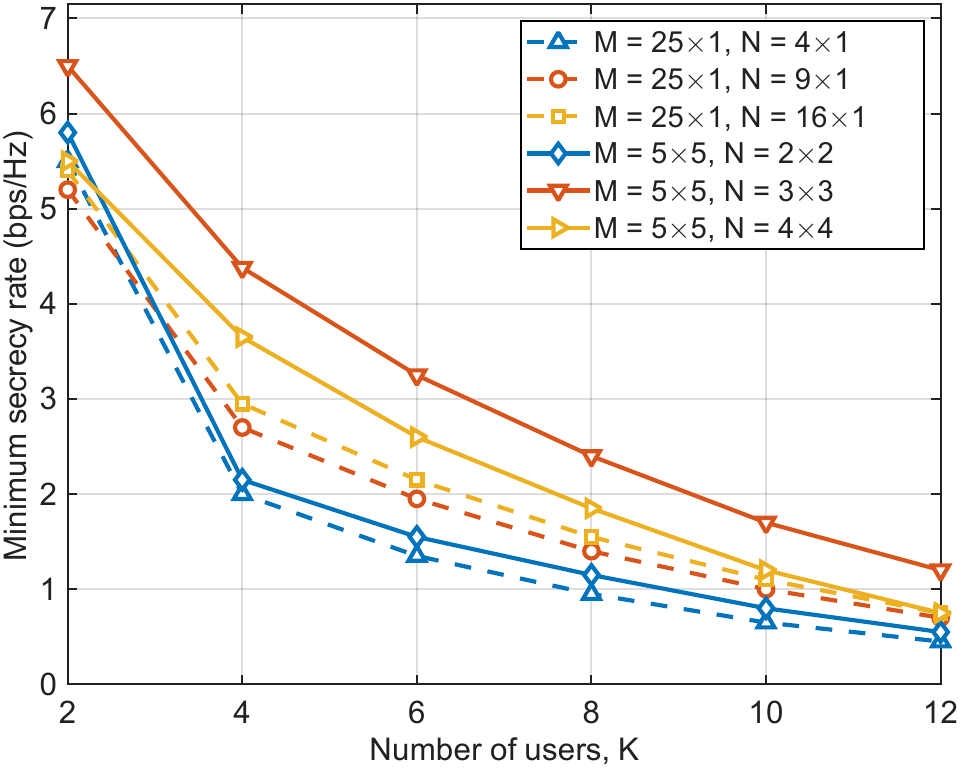}
\caption{Minimum secrecy rate versus the number of users $K$ for different 1D and 2D MRIS configurations ($J=4$).}
\label{fig8}
\end{figure}

Fig. \ref{fig8} illustrates the variation of the minimum secrecy rate as the number of users $K$ increases from 2 to 12 with $J=4$ and a fixed S1 size of $M=\text{25}$. As expected, the secrecy rate for all configurations decreases monotonically with $K$ due to aggravated multi-user interference and stricter QoS constraints. For a given number of S2, the 2D configurations consistently outperform their 1D counterparts. This is because the secrecy performance gain does not stem from a greater number of beam patterns but rather from a more balanced aperture in the 2D case, which improves angular coverage across both azimuth and elevation, thereby conditioning the effective channel to suppress user–Eve leakage more efficiently under S2 mobility. However, it is noteworthy that the configurations corresponding to $N=\text{4}$ exhibit the lowest secrecy rate among the tested options, and their performance gap is minimal compared to other cases. This behavior aligns with the results in Fig.  \ref{fig6} and  Fig. \ref{fig7}, further confirming the geometric sensitivity of the $N=\text{2}\times\text{2}$ sub-aperture configuration.

\begin{figure}[!t]
\centering
\includegraphics[width=2.8in,trim=0 0 0 0, clip]{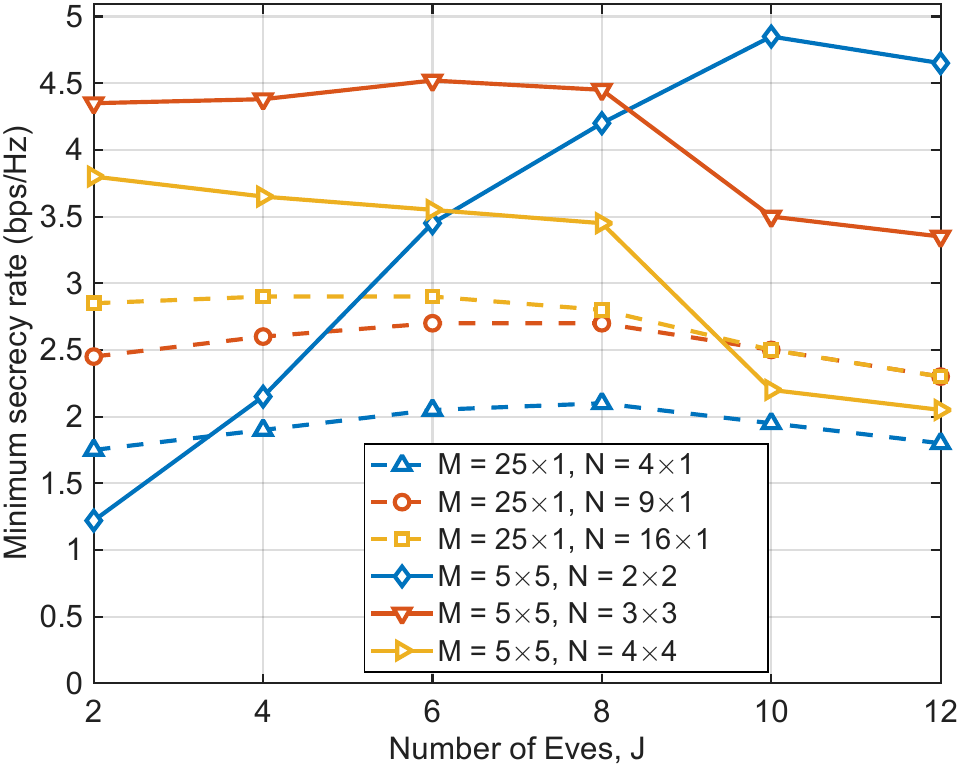}
\caption{Minimum secrecy rate versus the number of Eves $J$ for different 1D and 2D MRIS configurations ($K=4$).}
\label{fig9}
\vspace{-3mm}
\end{figure}

Fig. \ref{fig9} illustrates the variation of the minimum secrecy rate as the number of Eves $J$ increases from 2 to 12 with $K=4$ and a fixed S1 size of $M=\text{25}$. Unlike the monotonic degradation observed when scaling the number of users, the change in secrecy rate with respect to $J$ is more intricate, since adding Eve primarily tightens secrecy constraints rather than increasing multi-user interference. Overall, the 2D MRIS configurations achieve significantly higher secrecy rates than their 1D counterparts, and the $M=\text{5}\times\text{5}, N=\text{3}\times\text{3}$ layout maintains the best performance across the entire range of $J$. In contrast, the $N=\text{2}\times\text{2}$ configuration exhibits the lowest secrecy rate at
$J=$ 2, then increases substantially as $J$ increases to moderate values, and decreases again for larger $J$. This non-monotonic behavior is distinct from Fig. \ref{fig8}, where scaling $K$ leads to monotonic degradation. This indicates that the $N=\text{2}\times\text{2}$ layout shows better adaptability when more Eves are present, achieving higher performance at moderate $J$. Meanwhile, the $N=\text{3}\times\text{3}$ configuration maintains smoother and more stable performance across the entire range of $J$, demonstrating stronger robustness in complex spatial environments. 

\vspace{-0.5cm}
\section{Conclusion}
In this paper, we investigated a novel MRIS-assisted robust and secure ISAC system. A robust optimization framework was developed under Rician fading with imperfect CSI, where uncertainties in distance, angles, and NLoS fading were characterized by safe approximation bounds. The minimum secrecy rate of the system was maximized by jointly optimizing the BS transmit beamforming vectors, AN, MRIS phase shifts, and beam pattern scheduling. To solve the resulting non-convex problem, an efficient algorithm integrating AO and PDD frameworks was proposed. Numerical results demonstrated that the proposed design achieves significantly higher secrecy rates than the conventional SRIS baseline. Substantial secrecy gains were achieved by allocating or adding only a small number of elements to the movable sub-surface, confirming the cost-efficiency of the MRIS architecture. The results also revealed strong geometric sensitivity, where secrecy performance heavily depends on the spatial configuration of the two sub-surfaces and the spatial distribution of users and Eves. Certain MRIS layouts may induce phase-coupling effects, leading to unstable variations in secrecy rate across different spatial realizations. Future work may focus on developing adaptive MRIS control strategies to mitigate geometric sensitivity and enhance secrecy reliability. It is also important to examine trade-offs among mechanical mobility, energy efficiency, and secrecy robustness in practical ISAC systems. Furthermore, comparing MRIS with element-level MA-RIS can provide insight into how global surface motion and local element mobility affect secure transmission and overall robustness.

\vspace{-0.2cm}
\bibliographystyle{IEEEtran}
\bibliography{IEEEabrv,reference}

\vfill

\end{document}